\documentclass[manuscript]{aastex}
\usepackage{txfonts}
\usepackage{graphicx}

\slugcomment{Accepted to ApJ}

\shorttitle{Inflow motion in high-mass cores} \shortauthors{Wu et
al.}

\begin{document}

\title{Infall and outflow motions in the high-mass star forming complex G9.62+0.19}

\author{Tie Liu\altaffilmark{1}, Yuefang Wu\altaffilmark{1}, Sheng-Yuan Liu\altaffilmark{2}, Sheng-Li Qin\altaffilmark{3,4}, Yu-Nung Su\altaffilmark{2}, Huei-Ru Chen\altaffilmark{5,2} and Zhiyuan Ren\altaffilmark{1}}

\altaffiltext{1}{Department of Astronomy, Peking University, 100871,
Beijing China; liutiepku@gmail.com, ywu@pku.edu.cn }
\altaffiltext{2}{Institute of Astronomy and Astrophysics, Academia
Sinica, Taipei, Taiwan} \altaffiltext{3}{I. Physikalisches Institut,
Universit\"at zu K\"oln, Z\"ulpicher Str. 77, 50937 K\"oln, Germany}
\altaffiltext{4}{National Astronomical Observatories, Chinese
Academy of Sciences, Beijing, 100012} \altaffiltext{5}{Institute of
Astronomy and Department of Physics, National Tsing Hua University,
Hsinchu, Taiwan}

\begin{abstract}
We present the results of a high resolution study with the
Submillimeter Array towards the massive star forming complex
G9.62+0.19. Three sub-mm cores are detected in this region. The
masses are 13, 30 and 165 M$_{\sun}$ for the northern, middle and
southern dust cores, respectively. Infall motions are found with HCN
(4-3) and CS (7-6) lines at the middle core (G9.62+0.19 E). The
infall rate is $4.3\times10^{-3}~M_{\odot}\cdot$yr$^{-1}$. In the
southern core, a bipolar-outflow with a total mass about 26
M$_{\sun}$ and a mass-loss rate of
$3.6\times10^{-5}~M_{\odot}\cdot$yr$^{-1}$ is revealed in SO
($8_{7}-7_{7}$) line wing emission. CS (7-6) and HCN (4-3) lines
trace higher velocity gas than SO ($8_{7}-7_{7}$). G9.62+0.19 F is
confirmed to be the driving source of the outflow. We also analyze
the abundances of CS, SO and HCN along the redshifted outflow lobes.
The mass-velocity diagrams of the outflow lobes can be well fitted
by a single power law. The evolutionary sequence of the cm/mm cores
in this region are also analyzed. The results support that UC~H{\sc
ii} regions have a higher blue excess than their precursors.
\end{abstract}

\section{Introduction}

High-mass stars play a major role in the evolution of the Galaxy.
They are the principal sources of heavy elements and UV radiation
\citep{zin07}. However, the formation and evolution of high-mass
stars are still unclear. A possible evolution sequence of high-mass
stars from infrared dark clouds to classic H{\sc ii} regions has
been suggested \citep{van05}. But one of the major topics whether
high-mass stars form through accretion-disk-outflow, like low-mass
ones \citep{shu87}, or form via collision-coalescence
\citep{wol87,bon98} is still far from solved.

Yet more and more observations at various resolutions seem to
support the accretion-disk-outflow models rather than
collision-coalescence models. Disks are detected in several
high-mass star forming regions \citep{pat05,jia05,sri05}. Outflows
are found with a high detection rate as in low-mass cores in
single-dish surveys \citep{wu04,zha05,qin08a}. High resolution
studies have also confirmed that molecular outflows are common in
high-mass star forming regions
\citep{su04,qiu07,qin08b,qin08c,qiu09}. Searching for inflow motions
also has made large progress in recent years
\citep{wu03,ful05,wyr06,kla07,wu07,wu09,fur11}. Both infall and
outflow motions in the massive core JCMT 18354-0649S are detected
\citep{wu05}, and further confirmed by higher resolution
observations \citep{liu11}. Although accretion-disk-outflow systems
are found in high-mass star forming regions, there may be
differences between low- and high-mass formation.

The infall motion can be detected via "blue profile", a
double-peaked profile with the blueshifted peak being stronger for
optically thick lines and a single peak at the absorption part of
optically thick lines for optically thin lines, which is caused by
self absorption of the cooler outer infalling gas towards the warmer
central region \citep{zho93}. In contrast, the "red profile" where
the redshifted peak of a double-peaked profile being stronger for
optically thick lines is suggested as indicators for outflow
motions. \cite{mar97} defined the "blue excess" in a survey, E, as
E~=~(N$_{B}$-N$_{R}$)/N$_{T}$ (Mardones et al. 1997), where N$_{T}$
is number of sources, N$_{B}$ and N$_{R}$ mark the number of sources
with blue and red profiles, respectively. The blue excess seems to
be no significant differences among the low-mass cores in different
evolutionary phases. However, using the IRAM 30 m telescope,
\cite{wu07} found that UC~H{\sc ii} regions show a higher blue
excess than their precursors, indicating fundamental differences
between low- and high-mass-star forming conditions. The searches
need to be expanded.

Located at a distance of 5.7 kpc \citep{hof94}, G9.62+0.19 is a well
studied high-mass star forming region containing a cluster of H{\sc
ii} regions, which are probably at different evolutionary stages.
Multiwavelength VLA observations have identified nine radio
continuum sources (denoted from A-I) \citep{gar93,tes00}, and
components C-I are very compact ($<5\arcsec$ in diameter)
\citep{gar93,tes00}. As revealed in NH$_{3}$ (4,4), (5,5) and
CH$_{3}$CN (J=6-5), component F is a hot molecular core (HMC) and
hence likely the youngest source in the region
\citep{ces94,hof94,hof96b}. G9.62+0.19 E is a young massive star
surrounded by a very small UC~H{\sc ii} region and a dusty envelope
\citep{hof96b}, while G9.62+0.19 D a small cometary UC~H{\sc ii}
region excited by a B0.5 ZAMS star \citep{hof96b,tes00}. Both
G9.62+0.19 E and G9.62+0.19 D seem to be at a more evolved stage
than G9.62+0.19 F. Thus G9.62 complex is an ideal sample to examine
massive star forming activities including outflow and infall
motions.

Maser emissions of NH$_{3}$, H$_{2}$O, OH, and CH$_{3}$OH, as well
as the strong thermal NH$_{3}$ emissions were detected along a
narrow region with projected length $20\arcsec$ and
width$\leq2\arcsec$ \citep{hof94}. A possible explanation for this
alignment is compression of the molecular gas by shock front
originating from an even more evolved H{\sc ii} region to the west
of the star-forming front \citep{hof94}. High-velocity molecular
outflows also have been detected in this region, and G9.62+0.19 F is
believed to be the driving source \citep{gib04,hof01,su05}. However,
most of previous work was carried out at low frequencies, probing
low excitation conditions. To exam the hot dust/gas environment and
dynamical processes in this region, higher resolution studies at
high frequencies are needed. In this paper we report the results of
the Submillimeter Array (SMA\footnote {Submillimeter Array is a
joint project between the Smithsonian Astrophysical Observatory and
the Academia Sinica Institute of Astronomy and Astrophysics and is
funded by the Smithsonian Institution and the Academia Sinica.})
observations toward G9.62+0.19 region at 860 $\micron$.

\section{Observations}

The observations of G9.62+0.19 with the SMA were carried out on July
9th, 2005 with seven antennas in its compact configuration at 343
GHz for the lower sideband (LSB) and 353~GHz for the upper sideband
(USB). The T$_{sys}$ ranges from 210 to 990 K with a typical value
of 380 K at both sidebands during the observations. The observations
had two fields for the G9.62+0.19 complex to cover the entire region
with emissions. One phase reference center was RA(J2000)~=~18$^{\rm
h}$06$^{\rm m}$14.21$^{\rm s}$ and
DEC(J2000)~=~-$20\arcdeg31\arcmin46.2\arcsec$, and the other was
RA(J2000)~=~18$^{\rm h}$06$^{\rm m}$15.00$^{\rm s}$ and
DEC(J2000)~=~-$20\arcdeg31\arcmin34.20\arcsec$. Uranus and Neptune
were observed for antenna-based bandpass calibration. QSOs 1743-038
and 1911-201 were employed for antenna-based gain correction.
Neptune was used for flux-density calibration. The frequency spacing
across the spectra band was 0.8125~MHz, corresponding to a velocity
resolution of $\sim$0.7 km s$^{-1}$.

MIRIAD was employed for calibration and imaging \citep{sau95}. The
imaging was done to each field separately and the mosaic continuum
map was made using a linear mosaicing algorithm (task "linmos" in
MIRIAD). The 860 $\micron$ continuum data was acquired by averaging
over all the line-free channels in both sidebands. The spectral
cubes were constructed using the continuum-subtracted spectral
channels smoothed into a velocity resolution of 1 km~s$^{-1}$.
Additional self-calibration with models of the clean components from
previous imaging process was performed on the continuum data in
order to remove residual errors due to phase and amplitude problems,
and the gain solutions obtained from the continuum data were applied
to the line data. The synthesized beam size of the continuum
emission with robust weighting of 0.5 is
$2.76\arcsec\times1.88\arcsec$ (P.A.=$21.4\arcdeg$).

\section{Results}
\subsection{Continuum emission}
The 860 $\micron$ continuum image combining the visibility data from
both sidebands is shown in Figure 1 . Three sub-mm cores are
detected. The known cm and mm continuum components \citep{tes00} of
B, C, D, E, F, G, H, and I are marked by plus signs. Water masers
\citep{hof96a} are marked by open squares and methanol masers
\citep{nor93} by triangles. The near-IR sources
\citep{per03,tes98,linz05} are marked by filled circles. IRAC
sources are taken from the database of Galactic Legacy Infrared
Mid-Plane Survey Extraordinaire (GLIMPSE)
\footnote{http://irsa.ipac.caltech.edu/data/SPITZER/GLIMPSE/} and
labeled with asterisks. The northern core is located at south-east
of G9.62+0.19 C, and the middle core is associated with G9.62+0.19
E. The 860 $\micron$ continuum emission at the southern core is
concentrated on the hot molecular core G9.62+0.19 F and extends to
G9.62+0.19 D in the south and to G9.62+0.19 G in the north.

Gaussian fits were made to the continuum. The northern core seems to
be a point-like source. The middle core is very compact with a
deconvolved size of $\sim1.4\arcsec$. The southern core is found to
be elongated from north to south with an average size of
$2.4\arcsec$, containing at least three sources, D, F, and G. F is
at its peak position. The peak positions, sizes, peak intensities
and total fluxes of these three sub-mm cores are listed in Column
2-5 in Table 1. The physical properties of these cores will be
further discussed in section 4.2.

\subsection{Line emission}

Tens of molecular transitions including hot molecular lines
CH$_{3}$OH, HCOOCH$_{3}$, and CH$_{3}$OCH$_{3}$ are detected toward
both the middle and southern sub-mm cores, indicating these two
cores are hot and dense \citep{qin10}. Figure 2 presents the full
LSB and USB spectra in the UV domain over the shortest baseline. The
strongest lines are identified and labeled on the plots. Only HCN
(4-3) and CS (7-6) line emissions are detected towards the northern
core with our sensitivity. Thus we mainly focus on the middle and
southern cores in this paper. The systemic velocities of 2.1
km~s$^{-1}$ for the middle core and 5.2 km~s$^{-1}$ for the southern
core are obtained by averaging the V$_{lsr}$ of multiple singly
peaked lines. Six transitions of the thioformaldehyde (H$_{2}$CS)
and molecular transitions SO ($8_{7}$-7$_{7}$), CS (7-6), HC$^{15}$N
(4-3) and HCN (4-3) are analyzed here, while the others will be
discussed in another paper. We have made gaussian fits to the
beam-averaged spectra, and present the observed parameters of these
lines in Table 2.

\subsubsection{Line emission at the middle sub-mm core}

The integrated intensity maps of four transitions of H$_{2}$CS
towards the middle core are shown in the upper panels of Figure 3.
From (a) to (d), the upper level energy of H$_{2}$CS transitions
varies from $\sim90$~K to $\sim400$~K. The H$_{2}$CS emission is
spatially coincident with continuum emission of the middle core very
well. The Position-Velocity (P-V) diagram and first moment map of
H$_{2}$CS (10$_{2,8}$-9$_{2,7}$) emission are presented in Figure 4.
The P-V diagram is constructed across the peak of the continuum
along the N-S direction. From P-V diagram two emission peaks are
clearly revealed. The velocities of the two emission peaks are at 1
and 3 km~s$^{-1}$ with $1.5\arcsec$ spatial separation, indicating a
velocity gradient in N-S direction. The first moment map also shows
velocity changes in N-S direction. The small velocity gradient
detected in H$_{2}$CS (10$_{2,8}$-9$_{2,7}$) emission may indicate a
disk with a low inclination along the line of sight, which requires
further confirmation with higher angular resolution observations and
other molecular line tracers.

The spectra and integrated intensity maps of HC$^{15}$N (4-3) and SO
($8_{7}$-7$_{7}$) are presented in Figure 5. The two spectra seem to
be symmetric, and their cores are associated with that of the
continuum emission very well.

Figure 6 presents the spectra and P-V diagrams of HCN (4-3) and CS
(7-6) emissions of the middle core. HCN (4-3) and CS (7-6) show
asymmetric profile. The blue and red emission peaks of HCN (4-3) are
around 0 km~s$^{-1}$ and 6 km~s$^{-1}$, respectively. The
blueshifted emission of CS (7-6) peaks around 1 km~s$^{-1}$, while
the redshifted around 4 km~s$^{-1}$. We can see the blueshifted
emission of both HCN (4-3) and CS (3-2) is always stronger than the
redshifted emission and the absorption is also redshifted, which are
blue profiles (see Sect. 1). Besides the "blue profile", some weak
absorption dips are found around 10 km~s$^{-1}$ in both the spectra
and P-V diagrams of HCN (4-3) and CS (7-6), and further observations
are needed to determine the properties of these absorption dips. In
this paper we only pay attention to the "blue-profile" found in CS
(7-6) and HCN (4-3) emission.

The integrated intensity maps of HCN (4-3) and CS (7-6) towards the
middle core are presented in Figure 7. The HCN (4-3) and CS (7-6)
are associated with the dust emission.

\subsubsection{Line emission at the southern core}

The integrated intensity maps of four transitions of H$_{2}$CS at
the southern core are shown in the lower panels of Figure 3. The
upper level energy of H$_{2}$CS transitions varies from $\sim90$~K
to $\sim400$~K from panel (e) to panel (h). As the upper level
energy increases, the emission peak of the different transitions of
H$_{2}$CS moves from S-E to N-W, indicating a temperature gradient
in the southern core.

Averaged spectra of SO ($8_{7}$-7$_{7}$), HC$^{15}$N (4-3), HCN
(4-3) and CS (7-6) at the southern core are presented in Figure 8.
The spectra of SO ($8_{7}$-7$_{7}$) and HC$^{15}$N are averaged over
a region of 4$\arcsec$, while HCN (4-3) and CS (7-6) are averaged
over a region of 6$\arcsec$. SO ($8_{7}$-7$_{7}$) emission has a
total velocity extent of larger than 20 km~s$^{-1}$. From gaussian
fit to the spectrum, the peak velocity of SO emission is $5.1\pm0.1$
km~s$^{-1}$, coincident very well with the systemic velocity 5.2
km~s$^{-1}$. HC$^{15}$N (4-3) has a velocity extent of about 15
km~s$^{-1}$. The velocity extents of CS (7-6) and HCN (4-3) are as
high as 40 km~s$^{-1}$ and 60 km~s$^{-1}$, respectively. Emission
wings are clearly detected from the spectra of the four lines. A
"red-profile" is significantly exhibited in the spectra of CS (7-6)
and HCN (4-3), of which the redshifted emission is always stronger
than the blueshifted emission with an absorption dip at the
blueshifted side of the systemic velocity (5.2 km~s$^{-1}$). This
profile is caused by absorption of the colder blueshifted gas in
front of the hot core, indicating outflow motions. The "red-profile"
is consistent with that detected using single dish observations (see
Figure 6 of \cite{hof01}).

The integrated intensity maps of HC$^{15}$N (4-3) and SO
($8_{7}$-7$_{7}$) at the southern core are presented in Figure 9. To
avoid the influence of outflow motions, both the maps are integrated
from 2 km~s$^{-1}$ to 8 km~s$^{-1}$. Both the emission of HC$^{15}$N
(4-3) and SO ($8_{7}$-7$_{7}$) coincides with the cm/mm component F,
and extends from D to G.

As shown in the left panels of Figure 10, the high velocity gas of
HC$^{15}$N (4-3) and SO ($8_{7}$-7$_{7}$) can be identified by the
vertically dashed lines in the P-V diagrams. For SO
($8_{7}$-7$_{7}$), we integrate from -4 km~s$^{-1}$ $\leq$ V $\leq$
0 km~s$^{-1}$ for the blue wing and 10 km~s$^{-1}$ $\leq$ V $\leq$
14 km~s$^{-1}$ for the red wing, and present the contour map in (c)
of Figure 10. For HC$^{15}$N (4-3), only the red wing emission is
presented in (d) of Figure 10. The high velocity emission of both
HC$^{15}$N (4-3) and SO ($8_{7}$-7$_{7}$) is associated with core F,
indicating that core F is the driven source of the outflow. The blue
and red wings of SO ($8_{7}$-7$_{7}$) overlap to a large extent in
the contour maps, and hence the molecular outflow revealed by SO
($8_{7}$-7$_{7}$) is observed close to its flow axis.

Figure 11 presents the channel maps of CS (7-6) emission. The
redshifted high-velocity gas seems to be elongated from north-east
to south-west, while the blueshifted high-velocity gas from north to
south. The high velocity gas revealed by CS (7-6) is also very
obvious in the P-V diagram in Figure 13(d). As shown in the P-V
diagram, the blueshifted high-velocity gas extends about 8$\arcsec$
from north to south. The high-velocity emission integrated over the
wings (-12 km~s$^{-1}$ $\leq$ V $\leq$ -5 km~s$^{-1}$ for the blue
wing and 15 km~s$^{-1}$ $\leq$ V $\leq$ 22 km~s$^{-1}$ for the red
wing) is presented in Figure 13(e).

Figure 12 is the channel maps of HCN (4-3) emission. The maximum of
absorptions appears at around 0 km~s$^{-1}$. The redshifted
high-velocity gas seems to be elongated from west to east, while the
blueshifted high-velocity gas from north to south. At very high
velocity channels (V$\leq$ -16 km~s$^{-1}$), the blueshifted
emission is totally located at south-east. By comparing the channel
maps and P-V diagrams (see Figure 13) of HCN (4-3) and CS (7-6) at
velocity intervals -12 km~s$^{-1}$ $\leq$ V $\leq$ -5 km~s$^{-1}$
and 15 km~s$^{-1}$ $\leq$ V $\leq$ 22 km~s$^{-1}$, we find similar
structures in CS (7-6) and HCN (4-3) emissions. The high-velocity
emission of HCN (4-3) integrated from -12 km~s$^{-1}$ to -5
km~s$^{-1}$ for the blue wing and from 15 km~s$^{-1}$ to 22
km~s$^{-1}$ for the red wing is presented in the panel (b) of Figure
13. As of CS (7-6), the blueshifted gas revealed by HCN (4-3) is
elongated from north to south with the emission center located
between G9.62+0.19 F and G9.62+0.19 D, while the redshifted gas from
north-east to south-west. Two clumps are found in the blueshifted
high-velocity emission of HCN (4-3), which locate at north-west and
south-east of F, respectively.

In order to reveal the very high velocity emission traced by HCN
(4-3) but not CS (7-6), we integrate over the wings at much higher
velocities (-20 km~s$^{-1}$ $\leq$ V $\leq$ -13 km~s$^{-1}$ for the
blue wing and 23 km~s$^{-1}$ $\leq$ V $\leq$ 39 km~s$^{-1}$ for the
red wing), and present the integrated emission map in Figure 13(c).
The redshifted emission is elongated from north-east to south-west
with the emission center located between G9.62+0.19 F and G9.62+0.19
G, while the blueshifted emission center located between G9.62+0.19
F and G9.62+0.19 D. It is clearly seen that the high velocity gas
traced by SO ($8_{7}$-7$_{7}$), CS (7-6) and HCN (4-3) have
different spatial distributions, which should be caused by the
complicated interactions between the outflow and the ambient gas. It
may also indicate a change of the outflow axis. The change of
outflow axis is also found in IRAS 20126+4104 \citep{su07} and JCMT
18354-0649S \citep{liu11}.

From the integrated emission maps of SO ($8_{7}$-7$_{7}$), HCN (4-3)
and CS (7-6) high velocity gas, it is clearly seen that G9.62+0.19 F
is located at the middle of the redshifted and blueshifted lobes,
suggesting G9.62+0.19 F is the outflow driving source.

\section{Discussion}
\subsection{Rotational temperature of H$_{2}$CS transitions} Six transitions of
H$_{2}$CS have been detected in the middle and southern cores,
enabling us to estimate the rotational temperature. Under the
assumptions that the gas is optically thin under local thermodynamic
equilibrium and the gas emission fills the beam, the rotation
temperature and beam-averaged column density can be estimated using
the Rotational Temperature Diagram (RTD) by
\citep{cum86,tur91,liu02}
\begin{equation}
\textrm{ln}(\frac{N_{u}}{g_{u}}) =
\textrm{ln}(\frac{N_{T}}{Q_{rot}})-\frac{E_{u}}{T_{rot}} =
\textrm{ln}[2.04\times10^{20}\frac{\int~I(Jy~beam^{-1})dv(km~s^{-1})}{\theta_{a}\theta_{b}(arcsec^{2})g_{I}g_{K}\nu^{3}(GHz^{3})S\mu^{2}(debye^{2})}]
\end{equation}
where N$_{u}$ is the observed column density of the upper energy
level, g$_{u}$ is the degeneracy factor in the upper energy level,
N$_{T}$ is the total beam-averaged column density, Q$_{rot}$ is the
rotational partition function, E$_{u}$ is the upper level energy in
K, T$_{rot}$ is the rotation temperature, $\int$~I~dv is the
integrated intensity of the specific transition, $\theta_{a}$ and
$\theta_{b}$ are the FWHM beam size, g$_{K}$ is the K-ladder
degeneracy, g$_{I}$ is the degeneracy due to nuclear spin, $\nu$ is
the rest frequency, and S is line strength and $\mu$ the permanent
dipole moment. For H$_{2}$CS, the interchangeable nuclei are spin
$\frac{1}{2}$, leading to ortho- and para-forms with g$_{I}$
equaling $\frac{3}{4}$ and $\frac{1}{4}$, respectively
\citep{bl87,tur91}. The partition function Q$_{rot}$ of H$_{2}$CS is
\citep{bl87}
\begin{equation}
Q_{rot} = 2[\frac{\pi(kT_{rot})^{3}}{h^{3}ABC}]^{\frac{1}{2}}
\end{equation}
where k and h are the Boltzmann and Planck constants, respectively,
and A, B, and C are the rotation constants. Thus the rotation
temperature T$_{rot}$ and total column density N$_{T}$ can be
estimated by least-squares fitting to the multiple transitions. We
applied the RTD method towards D, E, F, G (see Figure 14), and the
fitting results are listed in the second and third columns of Table
3. The rotational temperature of the middle core (E) is 83$\pm$21 K.
In the southern core, the rotational temperature estimated decreases
from G (91 K) to F (83 K) and D (43 K), suggesting the temperature
gradient in the southern core. The total column density of H$_{2}$CS
ranges from 1.3$\times$10$^{15}$ (G) to
3.8$\times$10$^{15}$~cm$^{-2}$ (D).

However, the filling factor and the optical depth correction were
not taken account of in the RTD method. To investigate their effect
we applied the Population Diagram (PD) analysis
\citep{gol99,wang10}. In the PD analysis, we have
\begin{equation}
\textrm{ln}(\frac{\hat{N_{u}}}{g_{u}}) =
\textrm{ln}(\frac{N_{T}}{Q_{rot}})-\frac{E_{u}}{T_{rot}}+\textrm{ln}(f)-\textrm{ln}(\frac{\tau}{1-e^{-\tau}})
\end{equation}
where $\hat{N_{u}}$ is the inferred column density of the upper
energy level from the PD analysis, f is the source filling factor
and $\tau$ is the optical depth. The optical depth $\tau$ can be
expressed by \citep{re04}

\begin{equation}
\tau =
\frac{8\pi^{3}S\mu^{2}\nu}{3k\Delta\textrm{v}T_{rot}}\frac{N_{T}}{Q_{rot}}e^{-\frac{E_{u}}{T_{rot}}}
\end{equation}

where $\Delta$v is the FWHM line width. Under LTE, the upper-level
populations, $\hat{N_{u}}$, can be predicted according to the
right-hand side of Equation (3) for a given set of total column
density, N$_{T}$, rotational temperature, T$_{rot}$, and source
filling factor, f.  The expected $\hat{N_{u}}$ were evaluated for
the parameter space of T$_{rot}$ = 10-500 K, N$_{T}$ =
10$^{14}$-10$^{17}$ cm$^{-2}$, and f between 0.01 and 1.0. To
compare the observed $N_{u}$ and the inferred $\hat{N_{u}}$, we
calculate the $\chi^{2}$ as:

\begin{equation}
\chi^{2} = \sum(\frac{N_{u}-\hat{N_{u}}}{\delta~N_{u}})^{2}
\end{equation}

where $\delta~N_{u}$ is the 1 $\sigma$ error of observed upper-state
column density. Although the $\chi^{2}$ is a good representation of
the goodness of fit, the parameter set with the lowest $\chi^{2}$
may not actually represent physical parameters very well due to the
uncertainties of the observed data. In order to find a
representative parameter set, we compute a weighted mean and
standard deviation for all the parameters, with the weights being
the inverse of the $\chi^{2}$. All the parameter sets where the
inferred upper-level population $\hat{N_{u}}$ corresponds with the
observed upper-level population $N_{u}$ within 3 $\sigma$ are used
to compute the weighted means and standard deviations. The derived
rotational temperature, total column density and filling factor of
each component are list in the [3-5] columns of Table 3. The
inferred optical depths of each line transition are listed in the
last six columns of Table 3. The rotational temperatures of D, E, F,
G are estimated to be 42$\pm$34, 92$\pm$74, 51$\pm$23 and 105$\pm$37
K, respectively. A temperature gradient in the southern core is also
revealed as in the RTD method. The four components D, E, F, G has
similar total column densities as high as 4$\times10^{16}$
cm$^{-2}$, about an order of magnitude higher than those obtained
from RTD method, which are mainly due to the small source filling
factor ($<$0.5). The optical depths of H$_{2}$CS
(10$_{0,10}$-9$_{0,9}$) at the four components are all much larger
than one, while the other transitions are always optically thin
except H$_{2}$CS (10$_{2,9}$-9$_{2,8}$) line at G.

\subsection{Core properties}
In the optically thin case, the total dust and gas masses of the
three sub-mm cores can be obtained with the formula
$M=S_{\nu}D^{2}/\kappa_{\nu}RB_{\nu}(T_{d})$ \citep{hil83}, where
$S_{\nu}$ is the flux at 860 $\micron$, D is the distance, R=0.01 is
the mass ratio of dust to gas, and $\kappa_{\nu}$ is dust opacity
per unit dust mass. $B_{\nu}(T_d)$ is the Planck function at a dust
temperature of T$_{d}$. We assume that T$_{d}$ equals the rotational
temperature of H$_{2}$CS. For the northern core, since only CS (7-6)
(upper energy E$_{u}$ = 65.8 K) and HCN (4-3) (E$_{u}$ = 42.5 K)
exhibit strong emission lines, we assume T$_{d}$ to be 50 K.
Together with the measurements at centimeter and millimeter
wavelengths, \cite{su05} extrapolated the ionized gas emission at
mm/submm wavelengths, and found that the 0.85 mm continuum
associated with components D, E, and F are dominated by thermal dust
emission. They have derived opacity index $\beta$ of components E
and F to be 1.2, and 0.8, respectively. For the northern sub-mm
core, ${\beta}=1.5$ is assumed. Using the above dust opacity
indexes, we adopt $\kappa_{\nu}$=2.0, 1.8, and 1.5~cm$^{2}$g$^{-1}$
for the northern, middle and southern cores, respectively
\citep{ossen94}. At the distance of 5.7 kpc , we get the total dust
and gas masses for these three cores, and list all the parameters in
Table 1. The deduced masses for the northern, middle and southern
cores are 13, 30, 165 M$_{\sun}$, respectively. The column density
of H$_{2}$ are 1.2$\times10^{24}$ and 2.1$\times10^{24}$ cm$^{-2}$
for the middle and southern sub-mm cores, respectively.

\subsection{Infall properties in the middle core}
In the middle core, both CS(7-6) and HCN(4-3) emission exhibits
"blue profile" feature, indicating infall motions of the gas
envelope toward the central star
\citep{ket88,zho93,zha98,wu03,wu05,wu07,ful05,wyr07,sun08}. The
velocity difference (0.9 km~s$^{-1}$) between the absorption dip in
CS (7-6) spectrum (3 km~s$^{-1}$) and the systemic velocity (2.1
km~s$^{-1}$) is taken as the infall velocity $V_{in}$. Since both
HCN (4-3) and CS (7-6) emissions are not resolved towards the middle
core, we simply take the dust core size as the radius of the infall
region, which may underestimate the infall rate derived below. The
kinematic mass infall rate can be calculated using
dM/dt=$4{\pi}R_{in}^{2}nmV_{in}$. n=1.5$\times10^{7}$cm$^{-3}$ is
the number density of this dust core. Taking Helium into account,
the mean molecular mass m is 1.36 times of H$_{2}$ molecule mass.
The infall rate calculated is
$4.3\times10^{-3}~M_{\odot}\cdot$yr$^{-1}$. For comparison, the
$V_{in}$ from pure free-infall assumption is also derived with the
formula $V_{in}^{2}=2GM/R_{in}$. The pure free-infall velocity is
$V_{in}=3.6~$km~s$^{-1}$ and thus the "gravitational" mass infall
rate is $1.7\times10^{-2}~M_{\sun}\cdot$yr$^{-1}$, which is larger
than the kinematic infall rate.

\subsection{Outflow properties in the southern core}

\subsubsection{Shock chemistry in the outflow region of the southern core}

Observations have suggested that there are important differences in
molecular abundances in different outflow regions
\citep{bac97,cho04,jor04,cod05}. Significant abundance enhancements
are found in the shocked region for sulfur-bearing molecules
\citep{bac97,jor04}, and the abundance of HCN in outflow regions is
related to atomic carbon abundance \citep{cho02}. However, previous
studies of the chemical impact of outflows are confined to the well
collimated outflows around Class 0 sources, while such studies
especially high resolution studies on massive outflows are rare
\citep{bac97,jor04,arc07}.

A red and bright IRAC source is found to be associated with the
southern core. The magnitudes of the IRAC source at 3.6 $\micron$,
4.5 $\micron$ and 5.8 $\micron$ are $10.102\pm0.093$,
$8.361\pm0.108$ and $7.778\pm0.302$ mag, respectively. The [3.6-4.5]
color is as large as 1.74, indicating shocked emission in the
southern core \citep{tak10}. Maser emissions of NH$_{3}$, H$_{2}$O,
OH, and CH$_{3}$OH, as well as the strong thermal NH$_{3}$ emissions
also uncover the existence of the shocked gas \citep{hof94}.
Outflows can be revealed from shocked H$_{2}$ emission probed by the
strong and extended emission at the 4.5 $\micron$ band
\citep{qiu08,tak10}. Thus the massive outflow in the southern core
of G9.62 complex provides an ideal sample to study shock chemistry.

The fractional abundance of a certain molecule is defined as $\chi =
N_{T}/N_{H_{2}}$, where $N_{T}$ is the total column density of a
specific molecule and $N_{H_{2}}$ is the H$_{2}$ column density.
Assuming that the gas is optically thin and the emission fills the
beam, the beam-averaged total column density of a specific molecule
can be obtained from:

\begin{equation}
N_{T} =
2.04\times10^{20}\frac{\int~I(Jy~beam^{-1})dv(km~s^{-1})Q_{rot}e^{E_{u}/T_{rot}}}{\theta_{a}\theta_{b}(arcsec^{2})g_{I}g_{K}\nu^{3}(GHz^{3})S\mu^{2}(debye^{2})}
\end{equation}

Assuming that T$_{rot}$ of HC$^{15}$N equals to that of H$_{2}$CS
and the gas is optically thin, $N_{T}$ of HC$^{15}$N is calculated
to be $3.0\times10^{13}$~cm$^{-2}$ at the core region. At the
galactocentric distance of 3 kpc for G9.62+0.19 (Scoville et
al.1987,Hofner et al.1994), the abundance ratio
[$^{14}$N]/[$^{15}$N]$~\approx~350$ (Wilson and Rood. 1994). Thus
the total column density of HCN at the core region should be
$1.1\times10^{16}$~cm$^{-2}$. Therefore, the fractional abundance of
HCN relative to H$_{2}$ at the core region is $5.2\times10^{-9}$.
HCN appears to be greatly enhanced in the outflow regions of the
L1157 \citep{bac97}, while has similar abundances in the outflow
region and the ambient cloud of NGC 1333¨CIRAS 2A \citep{jor04}.
Owing to the lack of a direct estimation of the H$_{2}$ column
density towards the outflow region, the fractional abundance of HCN
in the outflow region is also assigned to $5.2\times10^{-9}$ in
calculating the outflow parameters. Since the HC$^{15}$N emission
traces outflowing gas at much lower velocity than HCN, perhaps HCN
could be more enhanced in the high velocity component. With the
possibility of higher opacity and the lack of direct H$_{2}$ column
density measurement, the derived fractional abundance perhaps is a
lower limit anyway. \cite{su07} estimate an HCN abundance of
$\sim1-2\times10^{-8}$ in the massive outflow lobes of IRAS
20126+4104, which is comparable to our estimation here.

Since the blueshifted outflow gas traced by CS (7-6) and HCN (4-3)
suffers self-absorption, the abundance ratios among SO
($8_{7}-7_{7}$), CS (7-6), and HCN (4-3) were inferred from the
beam-averaged spectra taken from the redshifted outflow lobe. The
abundance ratio as a function of flow velocity (the outflow velocity
relative to the systemic velocity) of [CS/SO] is obtained assuming
five different excitation temperatures in the left panel of Figure
15. It can be seen that the abundance ratio of [CS/SO] increases
with the excited temperature. At each excitation temperature, the
abundance ratio of [CS/SO] has lower values at flow velocities less
than 6~km~s$^{-1}$, and higher values when V$_{flow}$ larger than
8~km~s$^{-1}$, whereas the abundance ratio seems to be constant at
flow velocities between 6~km~s$^{-1}$ and 8~km~s$^{-1}$. There are
two reasons for the lower abundance ratio when
V$_{flow}~<$~6~km~s$^{-1}$: first, the flux missing of CS (7-6) due
to the interferometer is more serious than SO ($8_{7}-7_{7}$);
second, CS (7-6) may be more optically thick at lower flow
velocities than SO ($8_{7}-7_{7}$). As shown in the P-V diagrams,
the emission region of CS (7-6) is much larger than SO
($8_{7}-7_{7}$) at high velocities. The higher abundance ratio when
V$_{flow}~>$~8~km~s$^{-1}$ is due to the smaller filling factor of
SO ($8_{7}-7_{7}$) emission. We propose the mean observed value
between 6~km~s$^{-1}$ and 8~km~s$^{-1}$ can represent the actual
abundance ratio of [CS/SO]. Assuming a typical excitation
temperature of T$_{ex}$=30 K \citep{wu04}, the abundance ratio of
[CS/SO] at the redshifted lobe is inferred as 0.7. \cite{nil00} find
that the [SO/CS] abundance ratios are strongly enhanced in the Orion
A and NGC 2071 outflows where the [SO/CS] ratios are estimated to be
about 24 and 2.2, respectively. However, the [SO/CS] abundance ratio
in the outflow of G9.62+0.19 is found to be 1.4, much lower than
that found in Orion A outflow.

As shown in the right panel of Figure 15, the abundance ratio of
[CS/HCN] decreases linearly with the flow velocity. To avoid the
missing flux difficulty, the abundance ratio is calculated at high
flow velocities larger than 7 km~s$^{-1}$. The decreasing of the
abundance ratio with velocity is because that the emission region
traced by CS (7-6) is always smaller than HCN (4-3), leading to
smaller filling factor for CS (7-6), which can be verified easily by
comparing the channel maps between CS (7-6) in Figure 11 and HCN
(4-3) in Figure 12 at high velocities. We fitted the observed data
with a linear function, and adopted the value at flow velocity of 10
km~s$^{-1}$ as the actual abundance ratio of [CS/HCN] in the outflow
region, which is [CS/HCN]=1.2. Since HCN fractional abundance is
$5.2\times10^{-9}$, the fractional abundances of CS and SO are
deduced to be $6.2\times10^{-9}$ and $8.9\times10^{-9}$,
respectively.

\subsubsection{Properties of the bipolar-outflow traced by SO ($8_{7}-7_{7}$) emission}
The SO ($8_{7}-7_{7}$) emission in the southern core shows line
wings, suggesting outflow motions. From the integrated intensity map
in Figure 10(c), we find the outflow lobes revealed by SO
($8_{7}-7_{7}$) emission peak at different position with different
position angle compared with previously reported H$_{2}$S
($2_{2,0}-2_{1,1}$) \citep{gib04} and HCO$^{+}$ (1-0) data
\citep{hof01}. But in the same sense, the blue- and red-lobes
revealed by SO overlap to a large extent as well as HCO$^{+}$ (1-0)
and H$_{2}$S ($2_{2,0}-2_{1,1}$) data, consistent with the argument
of the outflow being viewed pole-on \citep{hof01}.

The total mass of each outflow lobe is given by:
\begin{equation}
M_{flow} =
1.04~\times~10^{-4}D^{2}\frac{Q_{rot}e^{E_{u}/T_{rot}}}{\chi\nu^{3}S\mu^{2}}\int\frac{\tau}{1-e^{-\tau}}S_{\nu}dv
\end{equation}
where M$_{flow}$, D, S$_{\nu}$, $\chi$, and $\tau$ are the outflow
gas mass in M$_{\sun}$, source distance in kpc, line flux density in
Jy, relative abundance to H$_{2}$, and optical depth. The other
parameters have the same units as in equation (1). The fractional
abundance of SO is taken as $8.9\times10^{-9}$ (see Sec.4.4.1).
Assuming an excitation temperature of 30 K and the outflowing gas is
optically thin, the inferred outflow masses are 13 M$_{\sun}$ for
each of red and blueshifted lobes. Thus, the momentum can be
calculated by $P=\sum$M(v)dv, and the energy by
$E=\sum{\frac{1}{2}}$M(v)v$^{2}$dv, where $v$ is the flow velocity.
The derived parameters are listed in Table 4. The momentum and
energy of the red lobe are 82 M$_{\sun}\cdot$km~s$^{-1}$ and
$5.4\times10^{45}$ erg. For the blue lobe, the momentum and energy
are calculated to be 86 M$_{\sun}\cdot$km~s$^{-1}$ and
$5.8\times10^{45}$ erg. The dynamical timescale t$_{dyn}$ is
estimated as R/V$_{char}$, where R ($\sim$ 0.06 pc) is adopted as
the mean size of the outflow lobes assuming a collimation factor of
unity, and V$_{char}$ ($\sim$ 5.5 km~s$^{-1}$) is assumed as the
mass weighted mean velocity. Thus, the dynamic timescale is
estimated to be $1\times10^{4}$ year, which may be underestimated
due to the uncertainty of the outflow scale. The mechanical
luminosity L, and the mass-loss rate $\dot{M}$ are calculated as
L=E/t, $\dot{M} = P/(tV_{w})$, where the wind velocity V$_{w}$ is
assumed to be 500 km~s$^{-1}$ \citep{lam95}. The mechanical
luminosity L and the total mass-loss rate are estimated to be 9.3
L$_{\sun}$ and $3.6\times10^{-5}$~M$_{\sun}$$\cdot$yr$^{-1}$,
respectively.

\subsubsection{Very high-velocity gas detected in CS (7-6) emission}

The CS (7-6) emission at the southern core shows "red-profile" with
wide wings. We take $6.2\times10^{-9}$ as the fractional abundance
of CS relative to H$_{2}$ along the outflow lobes. Assuming T$_{ex}$
= 30 K, we derive the parameters for the CS outflow (Table 4) with
the same method used for SO ($8_{7}-7_{7}$). The outflow masses at
very high velocities (v$_{flow}~>~$10~km~s$^{-1}$) are 3.7
M$_{\sun}$ and 5.5 M$_{\sun}$ for the blueshifted and redshifted
lobes, respectively. The momentum and energy of the blueshifted lobe
at very high velocities are calculated to be 47
M$_{\sun}\cdot$km~s$^{-1}$ and $6.0\times10^{45}$ erg. For the
redshifted lobe, the momentum and energy at extremely high
velocities are calculated to be 68 M$_{\sun}\cdot$km~s$^{-1}$ and
$8.7\times10^{45}$ erg, which are similar to the blueshifted lobe.

\subsubsection{Very high-velocity gas detected in HCN (4-3) emission}

As discussed before, HCN (4-3) has a velocity extent of at least 60
km~s$^{-1}$, which traces extremely high-velocity (EHV) gas.
Adopting an excited temperature of 30 K, and an HCN-to-H$_{2}$
abundance ratio of $5.2\times10^{-9}$, the parameters of the outflow
are calculated and listed in Table 4. The outflow mass at very high
velocities (v$_{flow}~>~$10~km~s$^{-1}$) are 5.2 M$_{\sun}$ and 17.6
M$_{\sun}$ for the blueshifted and redshifted lobes, respectively.
The momentum and energy of the blueshifted lobe at very high
velocities are 85 M$_{\sun}\cdot$km~s$^{-1}$ and $1.4\times10^{46}$
erg. For the redshifted lobe, the momentum and energy at very high
velocities are 294 M$_{\sun}\cdot$km~s$^{-1}$ and $5.5\times10^{46}$
erg, which are larger than the blueshifted lobe.

\subsubsection{Mass-Velocity diagrams}
A broken power law, $dM(v) / dv \propto v^{-\gamma}$ usually
exhibits in molecular outflows near young stellar objects
\citep{cha96,lad96,rid01,su04,qiu07,qiu09}. The slope, $\gamma$,
typically ranging from 1 to 3 at low outflow velocities, and often
steepens at velocities larger than 10 km~s$^{-1}$ --- with $\gamma$
as large as 10 in some cases \citep{arc07}. Assuming optically thin,
the mass-velocity diagrams of the outflow at the southern core of
G9.62+0.19 complex are shown in Figure 16. SO ($8_{7}-7_{7}$), CS
(7-6), HCN (4-3) results were all used in the mass spectra. We
calculate the outflow mass traced by CS (7-6) and HCN (4-3) from
V$_{flow}$ of 10 km~s$^{-1}$ to avoid the absorption of the spectra.
Instead of broken power law appearance, the mass-velocity diagram of
blueshifted lobe can be well fitted by a single power law with a
power indexes of $2.28\pm0.23$. The mass-velocity diagram of
redshifted lobe can be well fitted by a single power law with a
power indexes of $1.70\pm0.17$ even though the mass drops more
rapidly after 25 km~s$^{-1}$. As marked by the dashed ellipse in the
right panel, the outflow mass revealed by CS (7-6) is much lower
than that revealed by HCN (4-3) at very high velocities. Despite the
CS data, the mass-velocity diagram of redshifted lobe at velocities
smaller than 25 km~s$^{-1}$ can be fitted by a single power law with
a much smaller power indexes of $1.08\pm0.09$. However, no
significant slope changes are found in both the red- and
blue-shifted lobes of the outflow at the southern core, which are
very different from those previous works.

\subsection{Different evolutionary stages of the three dust cores}

The northern core has the smallest diameter and mass among the three
cores. It seems likely to be a point source after deconvolution. It
is located south of the nominal radio UC~H{\sc ii} region G9.62+0.19
C. In this region, eight near-IR sources are detected in a diffuse
near-IR nebulosity at the west of the radio emission peak
\citep{per03}. The reddest one c7 (18$^{\rm h}$06$^{\rm
m}$14.34$^{\rm s}$,-20$\arcdeg$31$\arcmin$25.0$\arcsec$) is located
within $1\arcsec$ of the radio peak, while the faintest one c8
(18$^{\rm h}$06$^{\rm m}$14.42$^{\rm
s}$,-20$\arcdeg$31$\arcmin$27.4$\arcsec$) seems to be associated
with the sub-mm core detected in SMA observation. Source c8 is too
faint to be detected even at H band and also shows no emission at
12.5 $\micron$. In contrast to the bright, rich molecular spectrum
forest in the middle and southern sub-mm cores, the northern sub-mm
core lacks strong molecular emissions. There is also no other early
star forming signature such as masers associated with it. Since it
is with near-IR emission and at the edge of the UC~H{\sc ii} region
G9.62+0.19 C, the northern core may be just a remnant core in the
envelope of UC~H{\sc ii} region G9.62+0.19 C, which needs further
observations.

The middle core is associated with the hyper-compact H{\sc ii}
region G9.62+0.19 E \citep{gar93,kur02}. OH, H$_{2}$O, and NH$_{3}$
(5,5) masers have been detected near the radio emission peak
\citep{for89,hof94,hof96a}. Periodic class II methanol masers are
also found in G9.62+0.19 E \citep{van09,goe05,nor93}. Methanol
masers are believed to be a good tracer of young massive star
forming regions at stages earlier than relatively evolved UC~H{\sc
ii} regions \citep{lon07}. No infrared source coincides with
G9.62+0.19 E \citep{per03}. Hot molecular CH$_{3}$CN lines are
detected in this region, and a kinematic temperature of T$_{k}$ =
108 K was obtained from CH$_{3}$CN emission with LVG model
\citep{hof96b}, which is coincident with the rotational temperature
(T$_{rot}$ = 92 K) obtained from H$_{2}$CS emission. A spectra
forest including hot molecular lines, such as CH$_{3}$OH, is
detected towards G9.62+0.20 E, suggesting this core is in a hot
phase. Infall motions are traced by CS (7-6) and HCN (4-3) lines,
indicating active star forming in this region.  All above suggest
that G9.62+0.20 E is forming a massive young star.

The 860 $\micron$ dust emission of the southern core peaks at
G9.62+0.19 F, and extends from north to south. A hump structure is
found to the southeast of the emission peak, indicating another
possible sub-mm core. The previously recognized mm/cm cores
(G9.62+0.19 D, G) are at the edges of the southern core. G9.62+0.19
G is a weak radio source \citep{tes00}, while G9.62+0.19 D is
consistent with an isothermal UC~H{\sc ii} region excited by a B0.5
star \citep{hof96b}. Weaker radio emission was found at core F.
H$_{2}$O and OH masers are found across the whole sub-mm core from
north to south \citep{for89,hof96a}. A near-IR source with large NIR
excess is found to be associated with G9.62+0.19 F
\citep{tes98,per03}. With higher resolution observations
\cite{linz05} found four near-IR objects (F1-F4) in this core. F4 is
with little emission at K band but becomes redder at longer
wavelengths, which seems to correspond to the bright IRAC source
with large excess at 4.5 $\micron$. This object is the dominating
and closest associated source of core F. Core F is also confirmed to
be the driving source of an active outflow. All of above imply that
G9.62+0.19 F is a very young massive star forming region.

\subsection{Blue excess in high-mass star forming regions}

Wu et al. (2007) found that UC~H{\sc ii} regions show a higher blue
excess than UC~H{\sc ii} precursors with the IRAM 30 m telescope.
\cite{wyr06} also detected large blue excess in UC~H{\sc ii}
regions. "Blue profile" was detected with CS (7-6) and HCN (4-3)
lines in UC~H{\sc ii} region G9.62+0.19 E, while "red profile" in
hot molecular core G9.62+0.19 F, which coincides with their
argument. The detection of infall signature in G9.62+0.19 E also
coincides the interpretation that material is still accreted during
the UC~H{\sc ii} phase \citep{wu07,keto02}. Around younger cores,
the outflow is more active and cold than UC~H{\sc ii} regions, which
leads to more "red profile". While in UC~H{\sc ii} regions, the
outflows become weak. The surrounding gas of UC~H{\sc ii} regions is
thermalized and the temperature gradient towards the central star is
more likely to cause "blue profile", which results in the higher
blue excess than UC~H{\sc ii} precursors.

\section{Summary}
We have observed the G9.62+0.19 complex with the Submillimeter Array
(SMA) both in the 860 $\micron$ continuum and molecular lines
emission. The main results of this study are as follows:

1. Dust continuum at 860 $\micron$ reveals three sub-mm cores in
G9.62+0.19 star forming complex. With H$_{2}$CS as the rotational
temperature prober, the temperatures of E and F are estimated to be
92$\pm$74 and 51$\pm$23 K, respectively. The mass calculated are 13,
30, and 165 M$_{\sun}$ for the northern, middle and southern core.

2. In the middle core, HCN (4-3) and CS (7-6) spectra exhibit infall
signature. The infall rate calculated is
$4.3\times10^{-3}$~M$_{\sun}\cdot$yr$^{-1}$. The detection of infall
signature in G9.62+0.19 E coincides the interpretation that material
is still accreted after the onset of the UC~H{\sc ii} phase
\citep{wu07}.

3. In the southern core, high-velocity gas is detected in SO
($8_{8}-7_{7}$), CS (7-6) and HCN (4-3) lines. A bipolar-outflow
with a total mass about 26 M$_{\sun}$ and a mass-loss rate of
$3.6\times10^{-5}$~M$_{\sun}\cdot$yr$^{-1}$ is revealed in SO
($8_{8}-7_{7}$) line wing emission. G9.62+0.19 F is confirmed to be
the driving source of the outflows in the southern sub-mm core. The
abundance ratios of [CS/SO] and [CS/HCN] in the outflow region are
found to be 0.7 and 1.2, respectively. The abundance ratio [CS/HCN]
decreases with the flow velocity, indicating smaller outflow regions
revealed by CS (7-6) than that revealed by HCN (4-3). The
mass-velocity diagrams of the blueshifted and redshifted outflow
lobes can be well fitted by a single power law. The power indexes
for the blueshifted and redshifted lobes are $2.28\pm0.23$ and
$1.70\pm0.17$. No significant slope changes are found in the
mass-velocity diagrams.

4. The evolutionary sequence of the cm/mm cores in this region are
also analyzed. The northern core may be just a remnant core in the
envelope of UC~H{\sc ii} region G9.62+0.19 C, which needs further
observations. The middle core (G9.62+0.19 E) is in a hyper-compact
H{\sc ii} region. Core G9.62+0.19 F is confirmed to be a hot
molecular core.

5. The detection of blue profiles at the hyper-compact H{\sc ii}
region E and the red profiles at the hot molecular core F supports
the results of single-dish observations that UC~H{\sc ii} regions
have a higher blue excess than their precursors.

\section*{Acknowledgment}
\begin{acknowledgements}
We are grateful to the SMA staff making the observations. This work
is funded by Grants of NSFC No 10733030 and 10873019.
\end{acknowledgements}

\begin{figure}
\includegraphics[angle=-90,scale=.50]{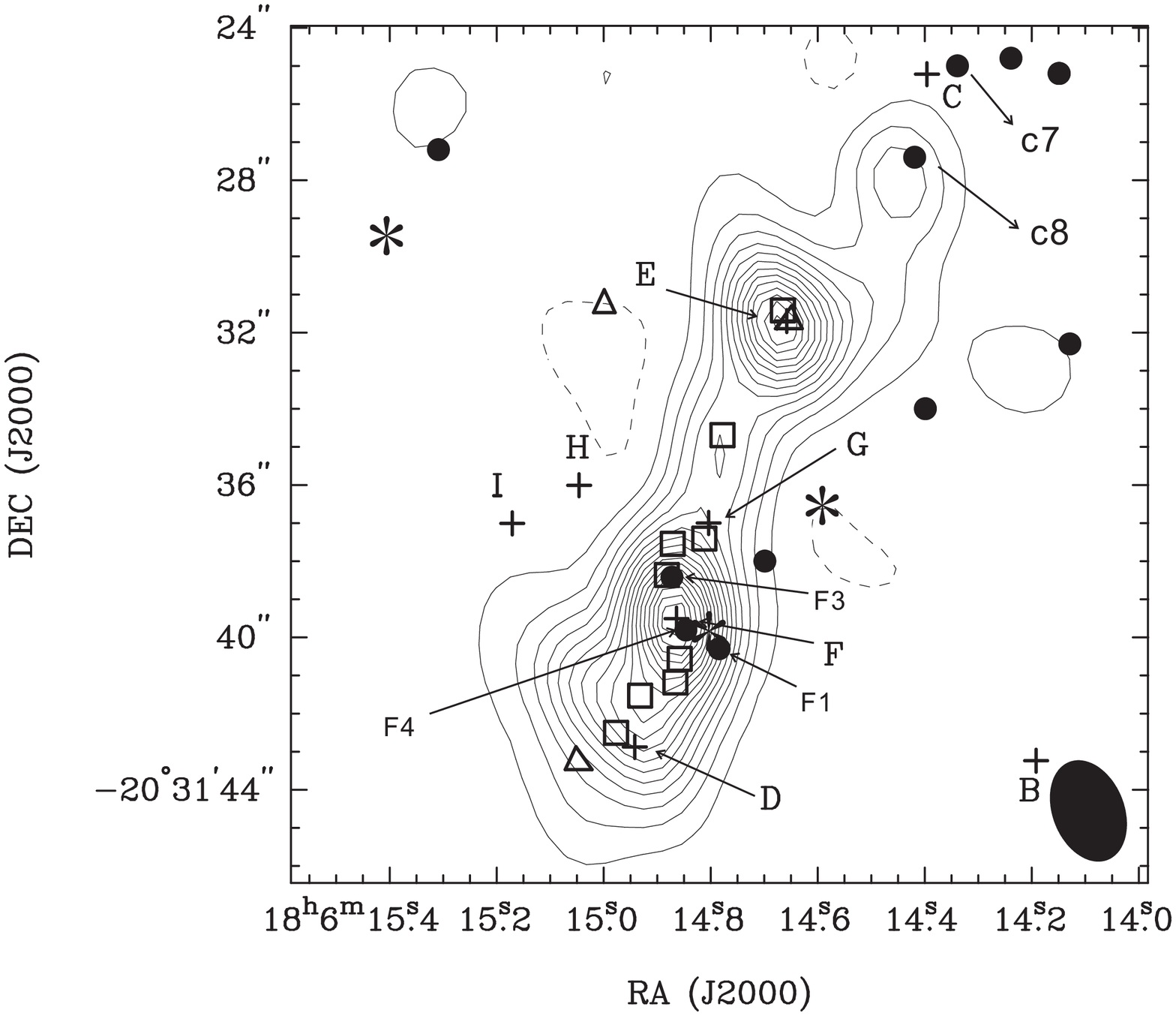}
\caption{The 860 $\micron$ continuum emission image. The contour
levels are from 0.03 Jy~beam$^{-1}$ (3$\sigma$) in steps of 0.06
Jy~beam$^{-1}$ (6$\sigma$). The known cm and mm continuum components
\citep{tes00} of B, C, D, E, F, G, H, and I are marked by plus
signs. Water masers \citep{hof96a} are marked by open squares and
methanol masers \citep{nor93} by triangles. The near-IR sources
\citep{per03,tes98,linz05} are marked by filled circles. IRAC
sources are marked with asterisks.}
\end{figure}

\begin{figure}
\includegraphics[angle=0,scale=.50]{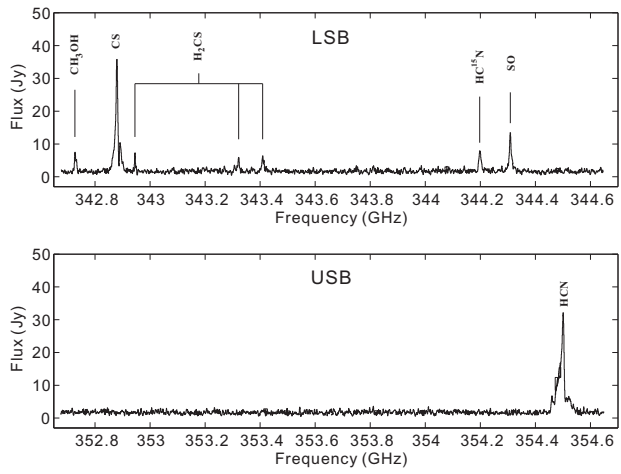}
\caption{The full LSB and USB spectra in the UV domain over the
shortest baseline. The strongest lines are identified and labeled on
the plots.}
\end{figure}

\begin{figure}
\includegraphics[angle=-90,scale=.50]{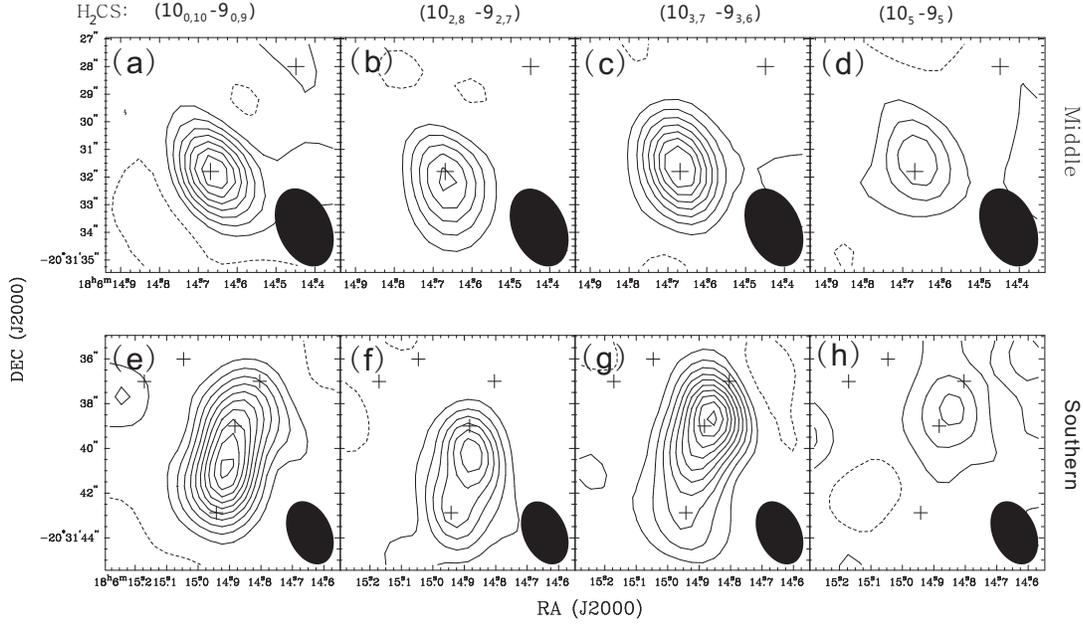}
\caption{Integrated intensity maps of four transitions of H$_{2}$CS
at the middle (upper panels) and southern cores (lower panels). The
known cm and mm continuum components are marked by plus signs as in
the continuum map. The contour levels in all the panels are from
3$\sigma$ in steps of 3$\sigma$. The rms levels are 0.3, 0.3, 0.3
and 0.2 Jy~beam$^{-1}\cdot$km~s$^{-1}$ for H$_{2}$CS
(10$_{0,10}$-9$_{0,9}$) in panels (a) and (e), H$_{2}$CS
(10$_{2,8}$-9$_{2,7}$) in panels (b) and (f), H$_{2}$CS
(10$_{3,7}$-9$_{3,6}$) in panels (c) and (g), and H$_{2}$CS
(10$_{5}$-9$_{5}$) in panels (d) and (h), respectively.}
\end{figure}

\begin{figure}
\includegraphics[angle=90,scale=.50]{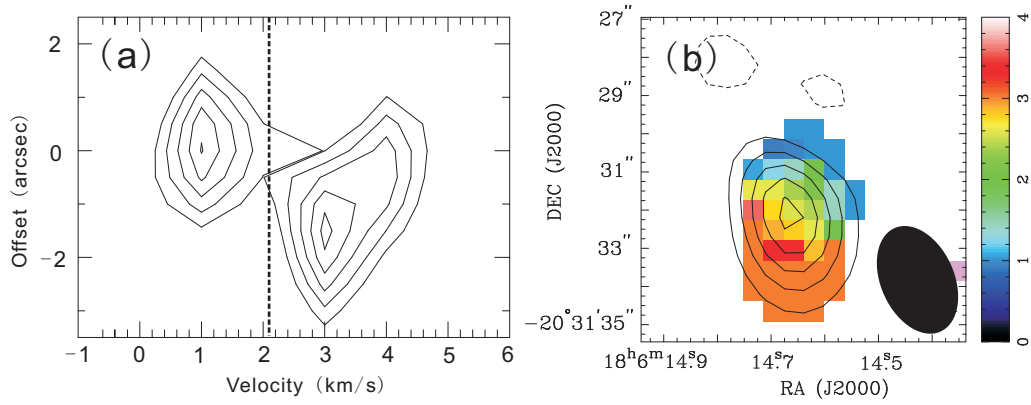}
\caption{The P-V diagram (left) and First moment map (right) of
H$_{2}$CS (10$_{2,8}$-9$_{2,7}$) emission at the middle core. (a)
The contours of the P-V diagram are from 0.6 to 1.4 in steps of 0.2
Jy~beam$^{-1}$ (1$\sigma$). (b) contour plot of H$_{2}$CS
(10$_{2,8}$-9$_{2,7}$) integrated intensity image overlayed on the
first moment map. The contours are from 0.9 (3$\sigma$) in steps of
0.9 Jy~beam$^{-1}\cdot$km~s$^{-1}$. The First moment map is
constructed from the data after imposing a cutoff of 3$\sigma$.}
\end{figure}

\begin{figure}
\includegraphics[angle=90,scale=.50]{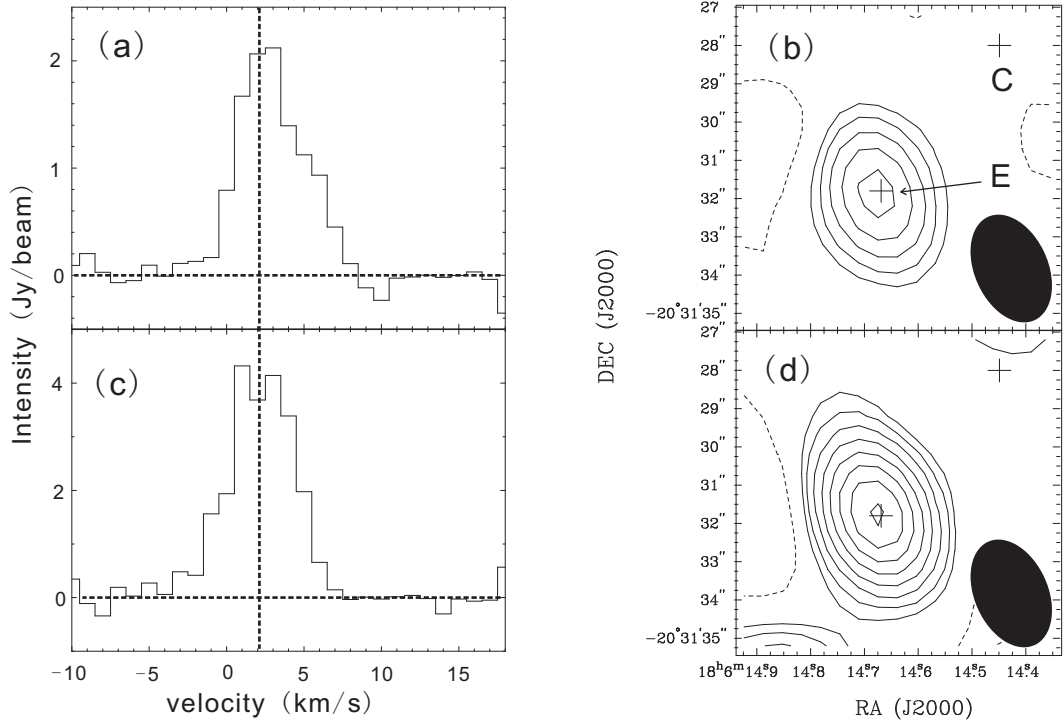}
\caption{Spectra and integrated intensity maps of HC$^{15}$N (4-3)
(upper panels) and SO ($8_{7}-7_{7}$) (lower panels) at the middle
core. The systemic velocity is marked with the thick vertical dashed
lines at the spectra panels. The known cm and mm continuum
components of C, and E are marked by plus signs at the integrated
maps as the continuum map. (a) the beam-averaged spectrum of
HC$^{15}$N (4-3) at E, (b) the integrated intensity map of
HC$^{15}$N (4-3). The contour levels are -1.2 (6$\sigma$), 1.2, 2.4,
4.2, 6.6, 9.6 Jy~beam$^{-1}\cdot$km~s$^{-1}$, (c) the beam-averaged
spectrum of SO ($8_{7}-7_{7}$) at E. (d) the integrated intensity
map of SO ($8_{7}-7_{7}$). The contour levels are -1.2 (6$\sigma$),
1.2, 2.4, 4.2, 6.6, 9.6, 13.2, 17.4, 22.2
Jy~beam$^{-1}\cdot$km~s$^{-1}$.}
\end{figure}

\begin{figure}
\includegraphics[angle=90,scale=.50]{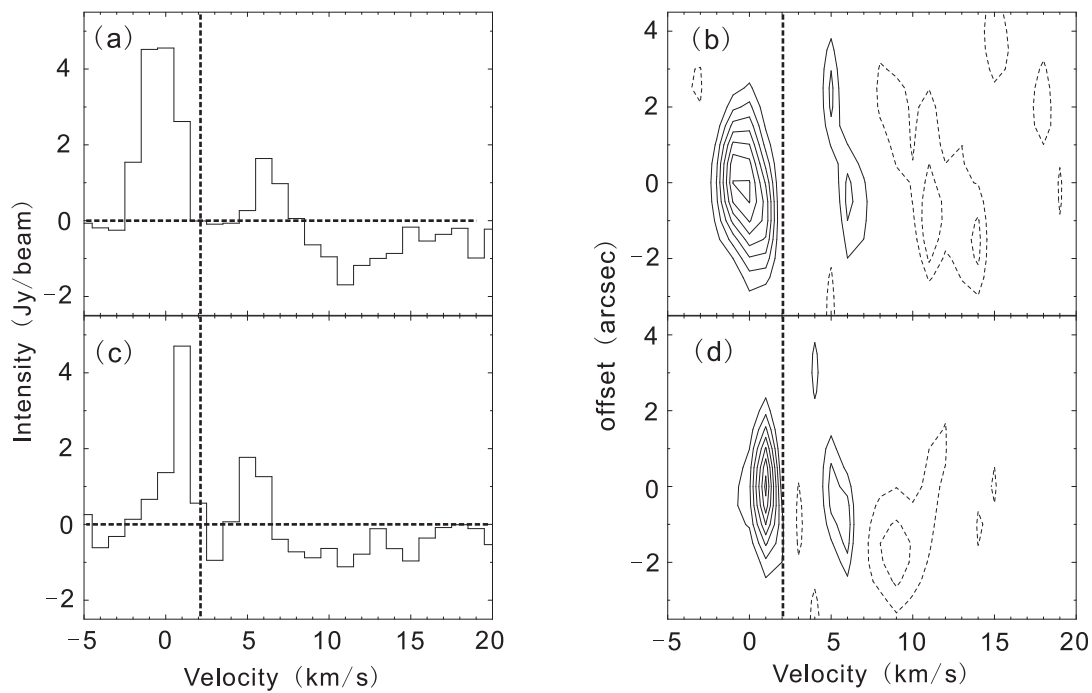}
\caption{Beam-averaged spectra and Position-Velocity (P-V) diagrams
of HCN (4-3) (upper panels) and CS (7-6) (lower panels) at the
middle core. The P-V diagrams are cut along a position angle of
0$\arcdeg$. (a) the beam-averaged spectrum of HCN (4-3) at E, (b)
the P-V diagram of HCN (4-3). The contour levels are -1.5
(5$\sigma$), -0.9, 0.9, 1.5, 2.1, 2.7, 3,3, 3.9, 4.5 Jy~beam$^{-1}$,
(c) the beam-averaged spectrum of CS (7-6) at E. (d) the P-V diagram
of CS (7-6). The contour levels are -1.5 (5$\sigma$), -0.9, 0.9,
1.5, 2.1, 2.7, 3,3, 3.9, 4.5 Jy~beam$^{-1}$.}
\end{figure}

\begin{figure}
\includegraphics[angle=-90,scale=.50]{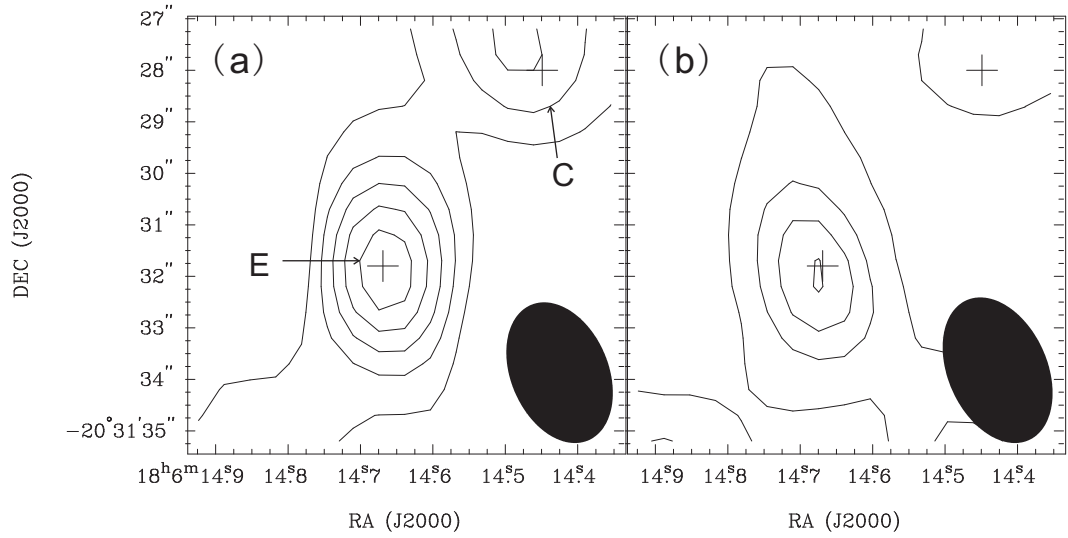}
\caption{Integrated intensity maps of HCN (4-3) (left) and CS (7-6)
(right) at the middle core. The contour levels in both maps are from
1.5 (5$\sigma$) in steps of 3 Jy~beam$^{-1}\cdot$km~s$^{-1}$. HCN
(4-3) is integrated from -3 to 7 km~s$^{-1}$, while CS (7-6) from -1
to 6 km~s$^{-1}$}
\end{figure}

\begin{figure}
\includegraphics[angle=90,scale=.50]{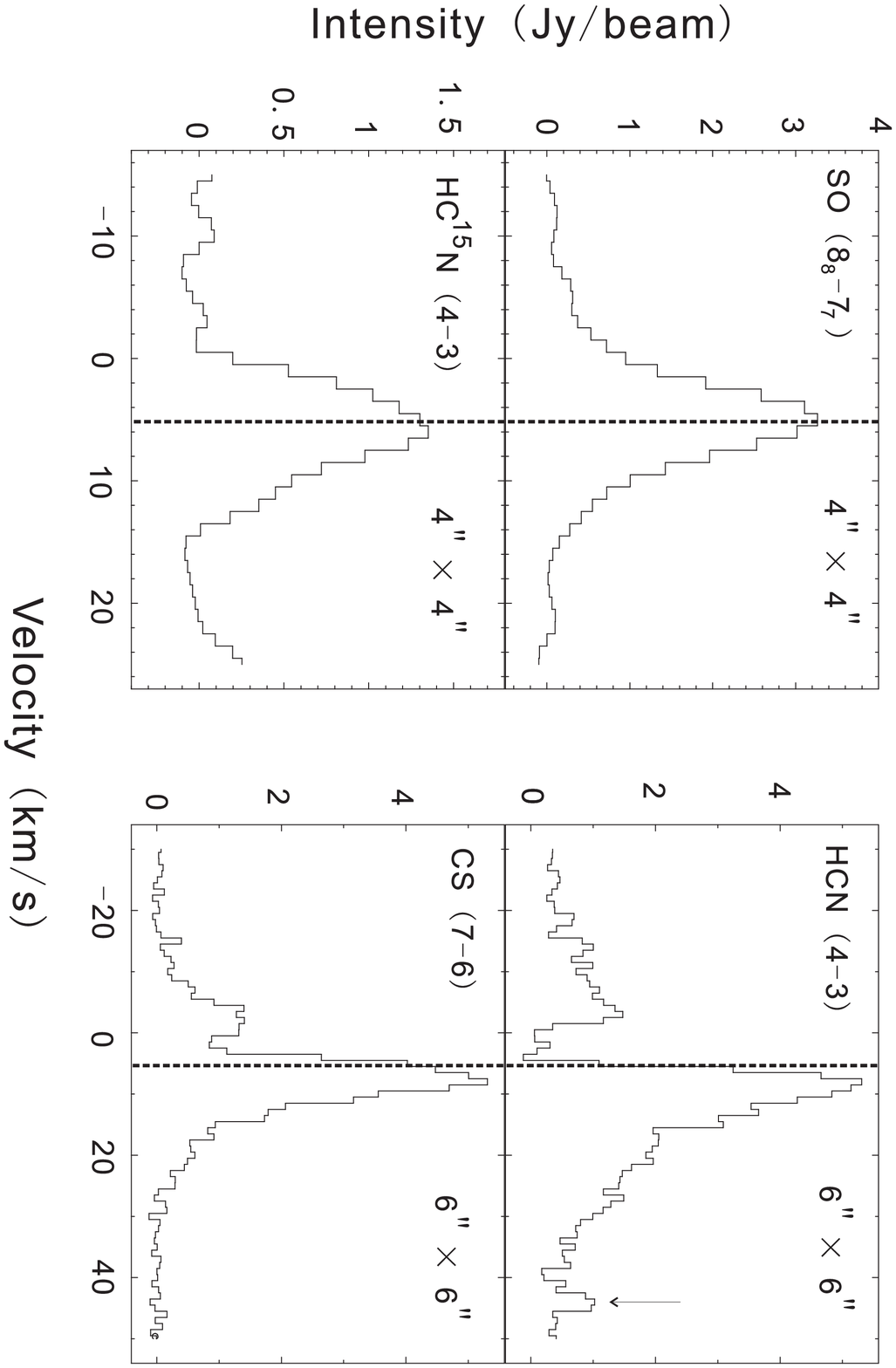}
\caption{Averaged spectra of SO ($8_{7}-7_{7}$) (upper-left),
HC$^{15}$N (4-3) (lower-left), HCN (4-3) (upper-right) and CS (7-6)
(lower-right) at the southern core. The spectra of SO
($8_{7}-7_{7}$) and HC$^{15}$N (4-3) are averaged over a region of
4$\arcsec$, while HCN (4-3) and CS (7-6) are averaged over a region
of 6$\arcsec$. HCN $\nu2=1$ (4-3) emission is marked by the arrow in
the upper-right panel, which can be clearly distinguished from the
red wing of HCN (4-3).}
\end{figure}

\begin{figure}
\includegraphics[angle=-90,scale=.50]{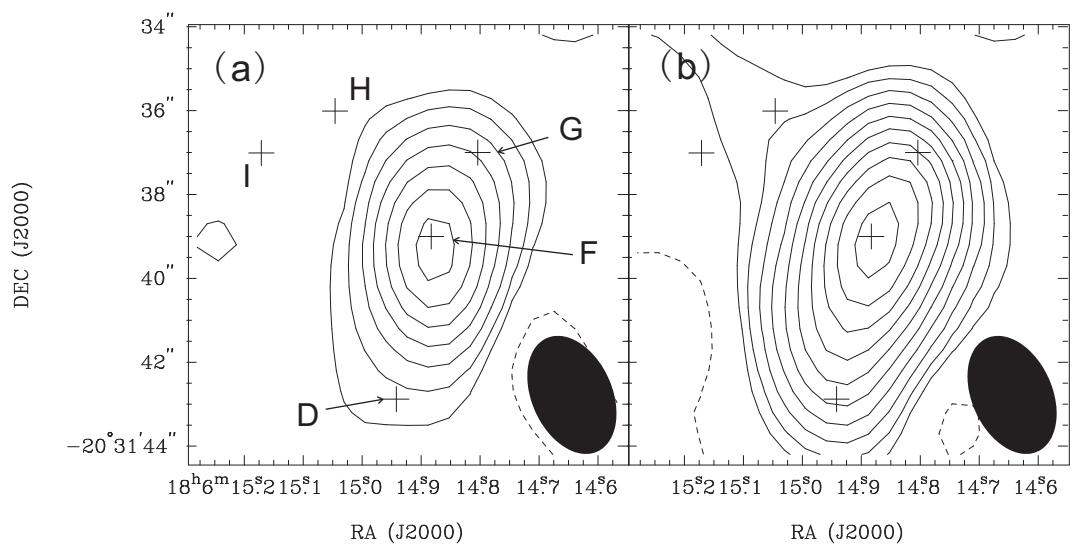}
\caption{The integrated intensity maps of HC$^{15}$N (4-3) (left
panel) and SO ($8_{7}-7_{7}$) (right panel) at the southern core. To
avoid the influence of outflow motions, both the maps are integrated
from 2 km~s$^{-1}$ to 8 km~s$^{-1}$. The contour levels are (a) -1.2
(6$\sigma$), 1.2, 2.4, 4.2, 6.6, 9.6, 13.2, 17.4
Jy~beam$^{-1}\cdot$km~s$^{-1}$ for HC$^{15}$N (4-3), (b) -1.2
(6$\sigma$), 1.2, 2.4, 4.2, 6.6, 9.6, 13.2, 17.4, 22.2, 27.6, 33.6
Jy~beam$^{-1}\cdot$km~s$^{-1}$ for SO ($8_{7}-7_{7}$)}
\end{figure}

\begin{figure}
\includegraphics[angle=-90,scale=.50]{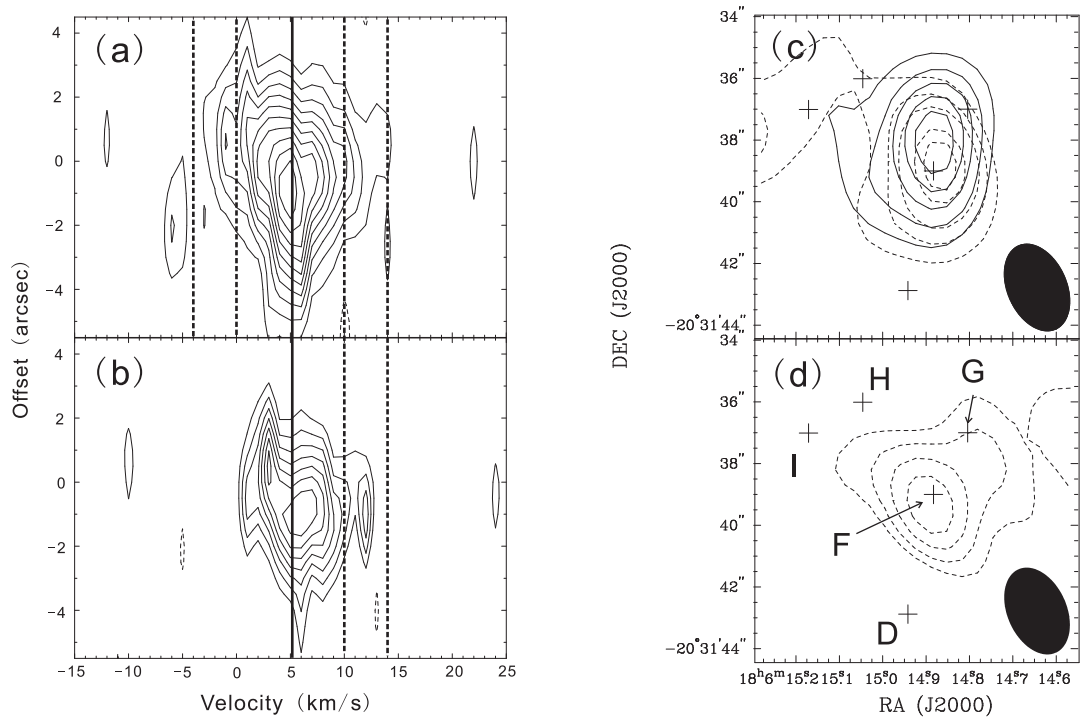}
\caption{P-V diagrams and integrated intensity maps of SO
($8_{7}-7_{7}$) (upper panels), and HC$^{15}$N (4-3) (lower panels)
at the southern core. The P-V diagrams are cut along N-S direction.
The vertical solid line in P-V diagrams labels the systemic
velocity.  The dashed and solid contours in the right panels show
the red- and blue-shifted emission, respectively. The integral
velocity intervals are marked by thick dashed lines in the P-V
diagrams. For both SO ($8_{7}-7_{7}$) and HC$^{15}$N (4-3), the
blue-shifted emission is integrated from -4 km~s$^{-1}$ to 0
km~s$^{-1}$, while the red-shifted emission from 10 km~s$^{-1}$ to
14 km~s$^{-1}$ in the integrated intensity maps. (a) P-V diagram of
SO ($8_{7}-7_{7}$). The contours are from 0.6 (3$\sigma$) in steps
of 0.6 Jy~beam$^{-1}$. (b)  P-V diagram of HC$^{15}$N (4-3). The
contours are from 0.6 (3$\sigma$) in steps of 0.4 Jy~beam$^{-1}$.
(c) Integrated intensity maps of SO ($8_{7}-7_{7}$) at line wings.
The contours are from 1 (5$\sigma$) in steps of 1
Jy~beam$^{-1}\cdot$km~s$^{-1}$ for both red- and blue-shifted
emission. (d) Integrated intensity maps of HC$^{15}$N (4-3) at red
wing. The contours are 0.6 (3$\sigma$), 1.2, 2, 3
Jy~beam$^{-1}\cdot$km~s$^{-1}$.}
\end{figure}

\begin{figure}
\includegraphics[angle=-90,scale=.50]{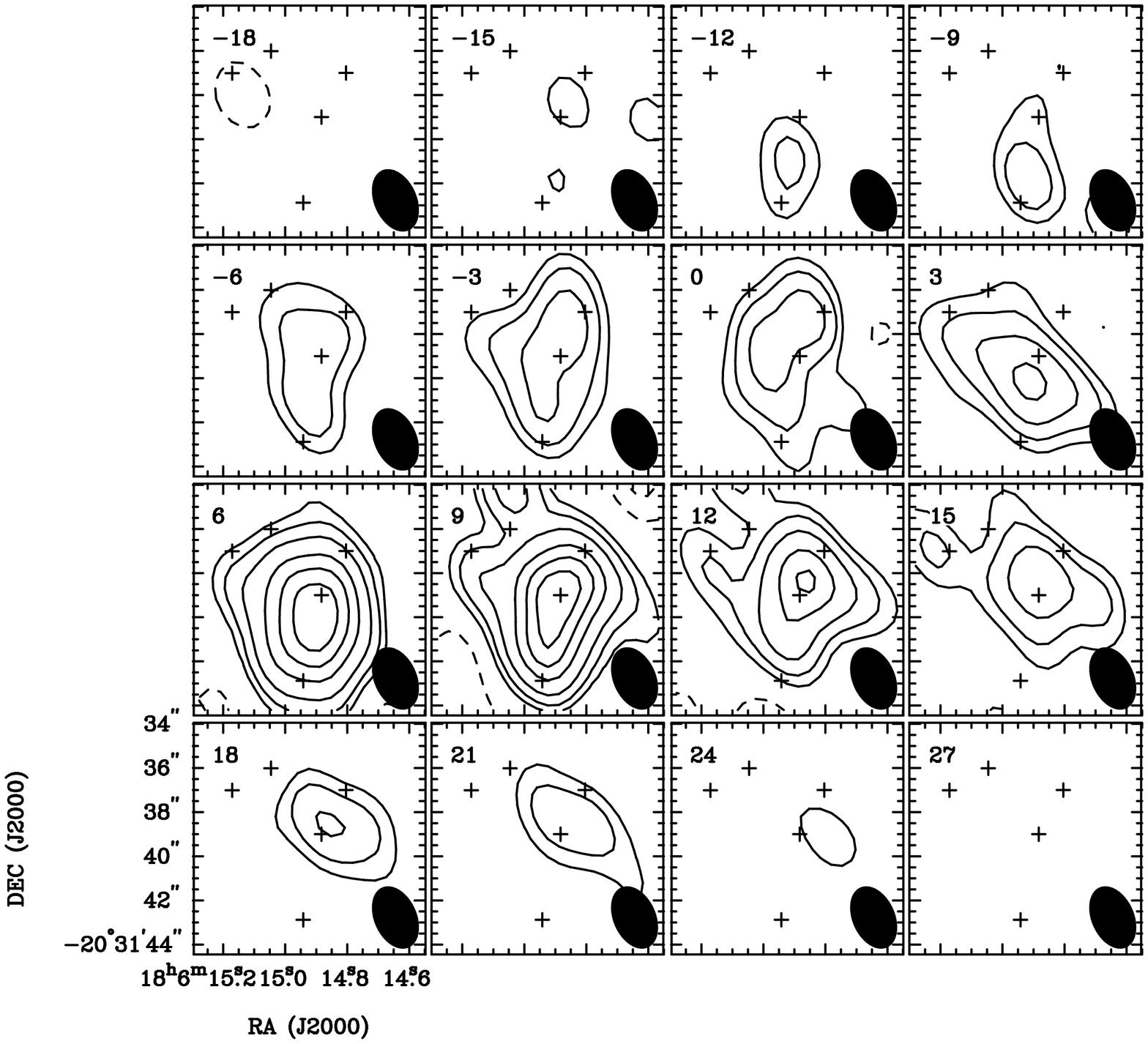}
\caption{CS (7-6) channel maps at the southern core, which is
smoothed to a velocity resolution of 3 km~s$^{-1}$. The contours are
-0.6 (3$\sigma$), 0.6, 1.2, 2.4, 4.8, 7.2, 9.6 Jy~beam$^{-1}$.}
\end{figure}

\begin{figure}
\includegraphics[angle=-90,scale=.50]{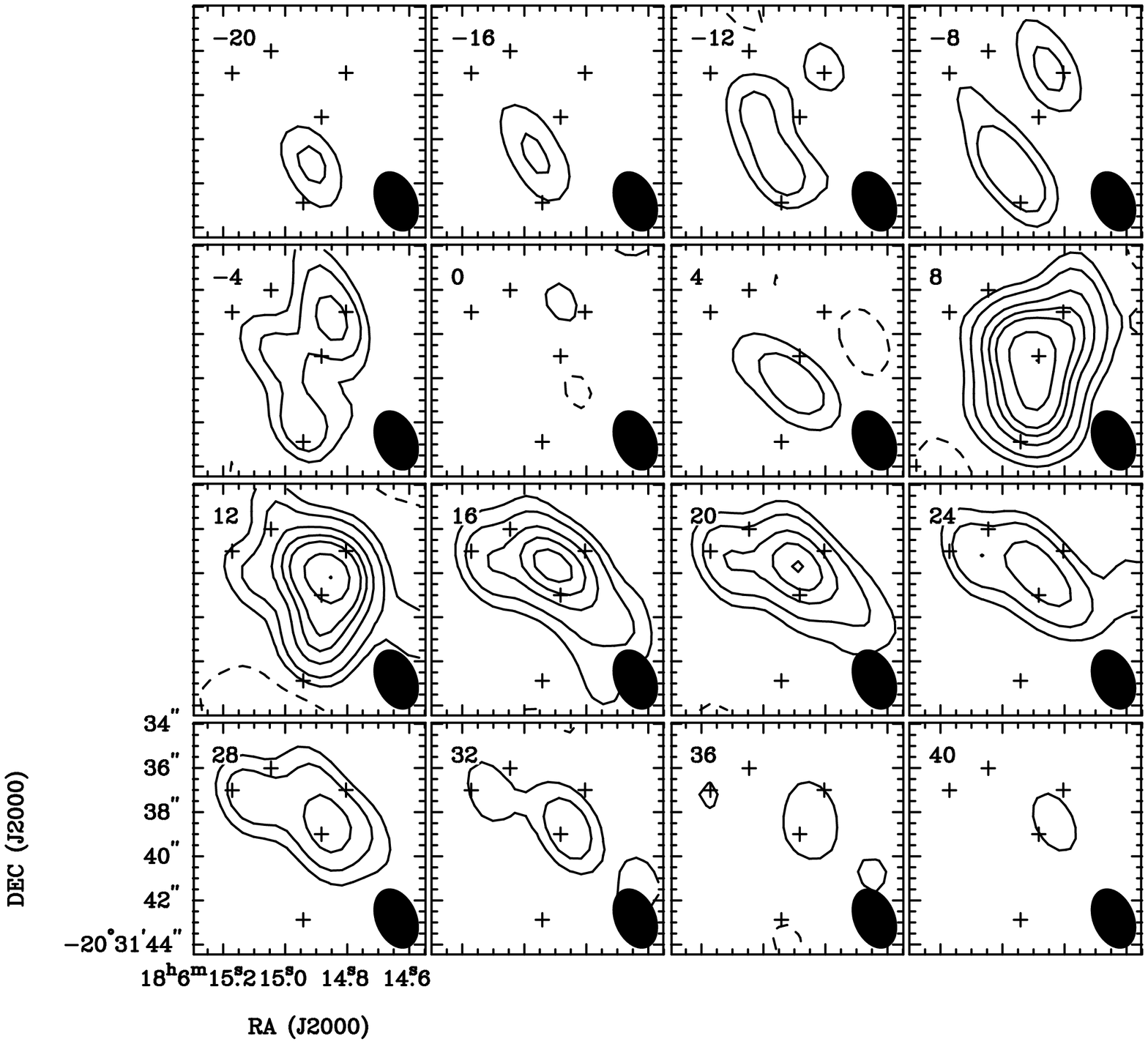}
\caption{HCN (4-3) channel maps at the southern core, which is
smoothed to a velocity resolution of 4 km~s$^{-1}$. The contours are
-0.6 (3$\sigma$), 0.6, 1.2, 2.4, 3.6, 4.8, 7.2 Jy~beam$^{-1}$.}
\end{figure}

\begin{figure}
\includegraphics[angle=-90,scale=.50]{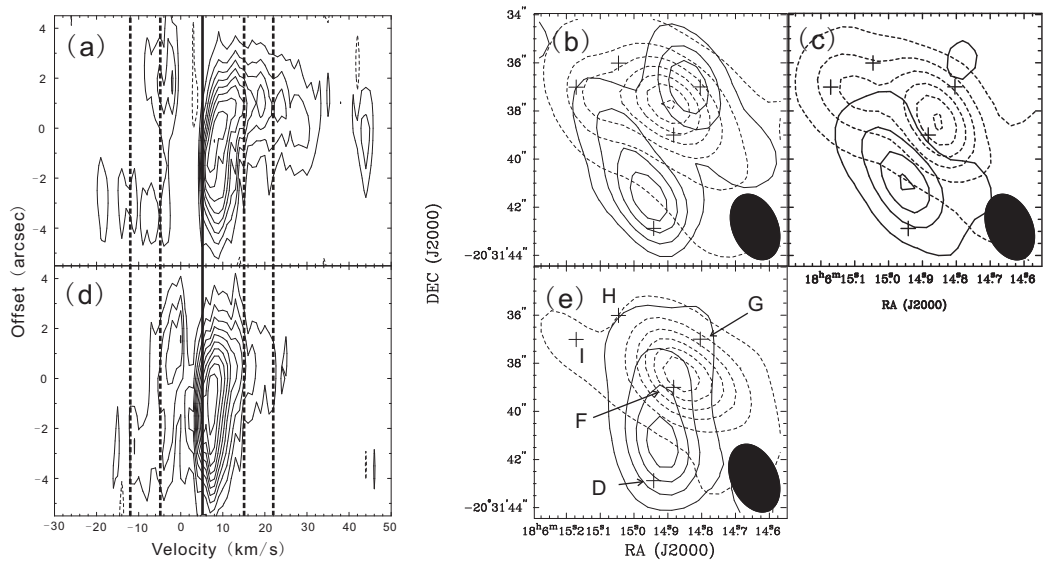}
\caption{P-V diagrams and integrated intensity maps of HCN (4-3)
(upper panels), and CS (7-6) (lower panels) at the southern core.
The P-V diagrams are cut along N-S direction. The vertical solid
line in P-V diagrams labels the systemic velocity. The dashed and
solid contours in the right panels show the red- and blue-shifted
emission, respectively. The blue- and red-shifted emission in the
integrated maps are integrated from -12 km~s$^{-1}$ to -5
km~s$^{-1}$ and 15 km~s$^{-1}$ to 22 km~s$^{-1}$, respectively in
(b) and (e) panels. (a) P-V diagram of HCN (4-3). The contours are
from 0.9 (3$\sigma$) in steps of 1.2 Jy~beam$^{-1}$. (b) Integrated
intensity maps of HCN (4-3) at line wings. The contours are 1.5
(5$\sigma$), 4.5, 7.5, 10.5 Jy~beam$^{-1}\cdot$km~s$^{-1}$. (c) The
integrated intensity maps of HCN (4-3) at extremely high velocities.
The blue- and red-shifted emission in the integrated maps are
integrated from -20 km~s$^{-1}$ to -13 km~s$^{-1}$ and 23
km~s$^{-1}$ to 39 km~s$^{-1}$, respectively. The contours are from
1.5 (5$\sigma$) in steps of 3 Jy~beam$^{-1}\cdot$km~s$^{-1}$ for
both blue- and red-shifted emission. (d)  P-V diagram of CS (7-6).
The contours are from 0.9 (3$\sigma$) in steps of 1.2
Jy~beam$^{-1}$. (e) Integrated intensity maps of CS (7-6) at line
wings. The contours are 1.5 (5$\sigma$), 4.5, 7.5, 10.5
Jy~beam$^{-1}\cdot$km~s$^{-1}$ }
\end{figure}

\begin{figure}
\begin{minipage}[c]{0.5\textwidth}
  \centering
  \includegraphics[width=80mm,height=70mm,angle=0]{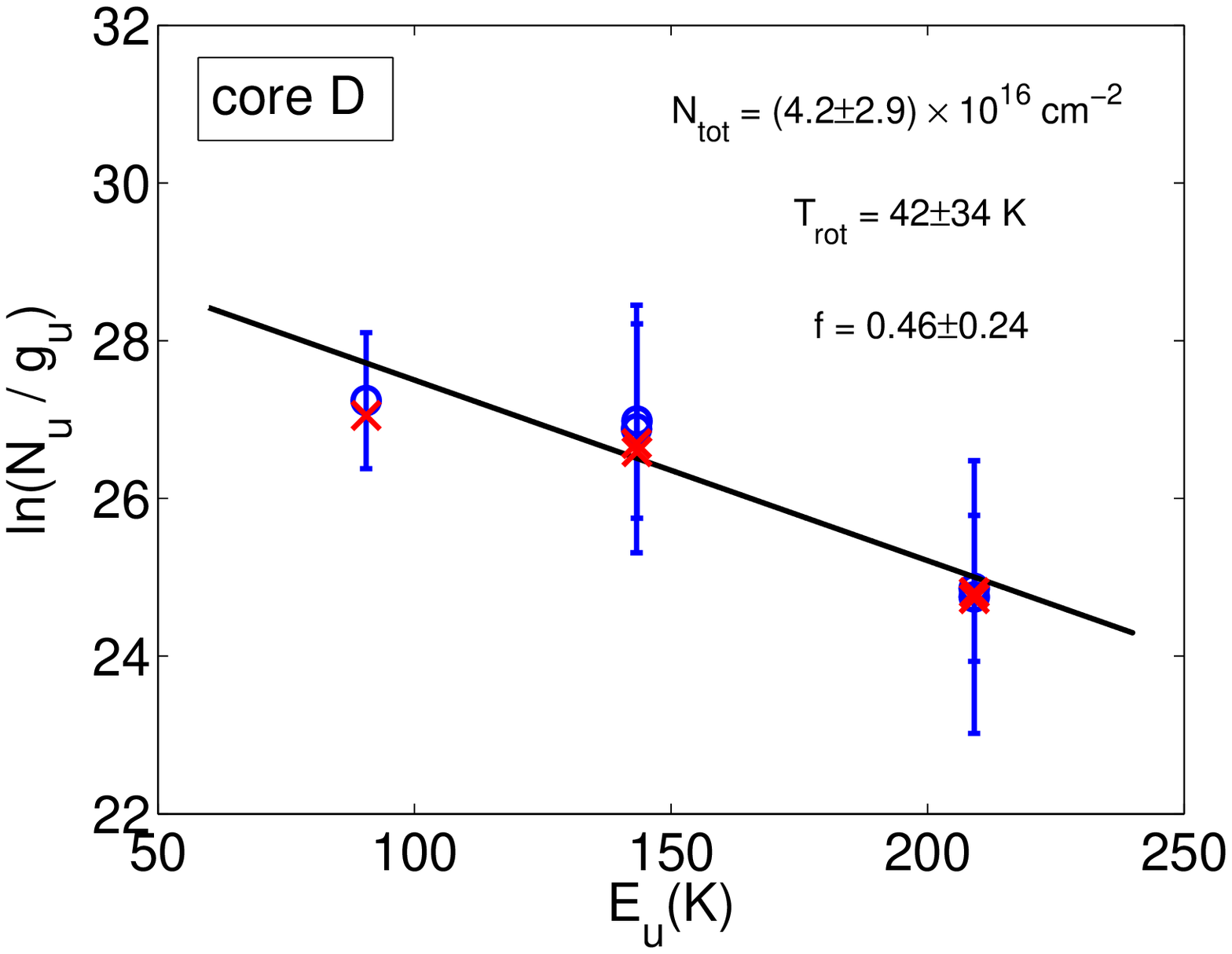}
\end{minipage}
\begin{minipage}[c]{0.5\textwidth}
  \centering
  \includegraphics[width=80mm,height=70mm,angle=0]{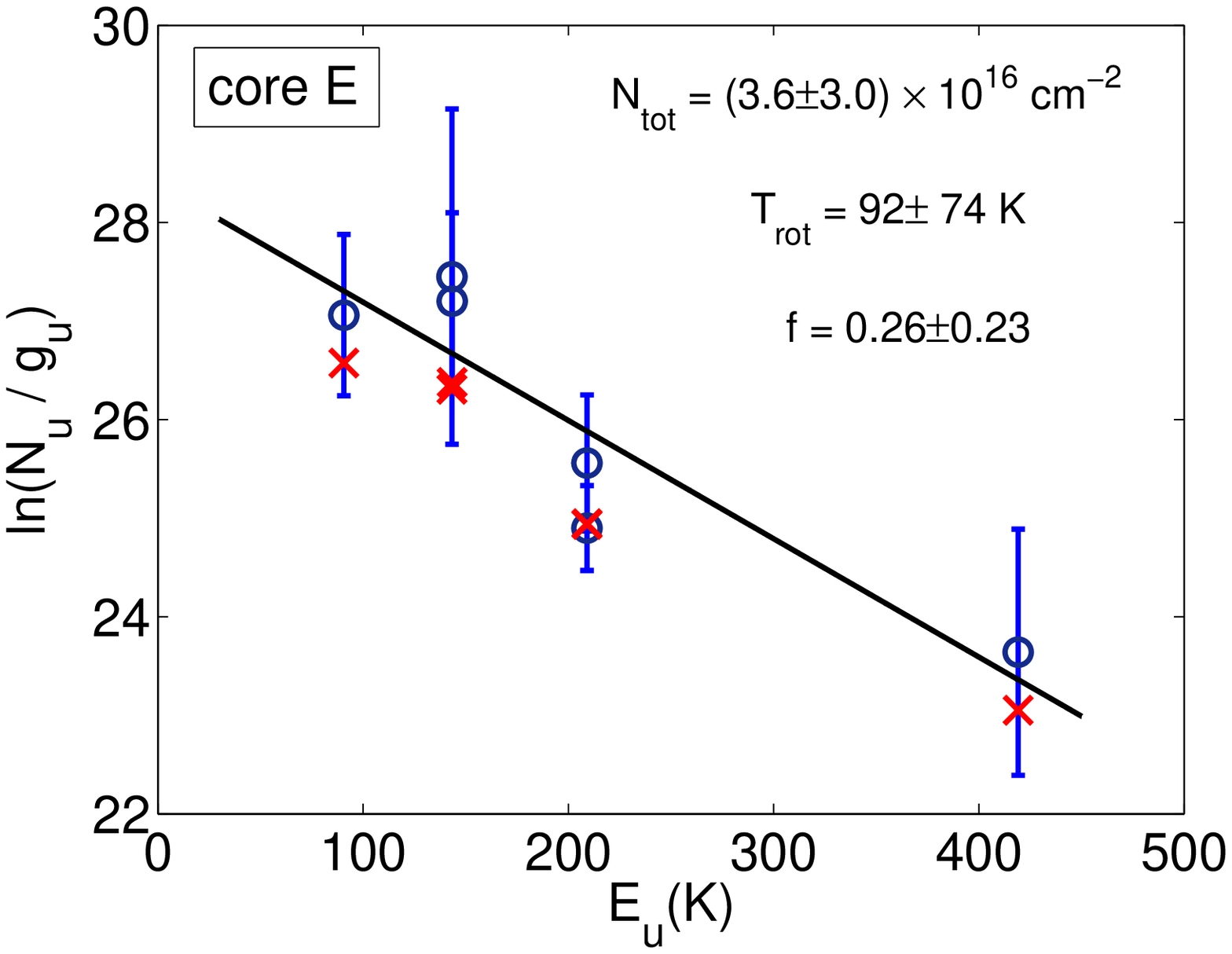}
\end{minipage}
\begin{minipage}[c]{0.5\textwidth}
  \centering
  \includegraphics[width=80mm,height=70mm,angle=0]{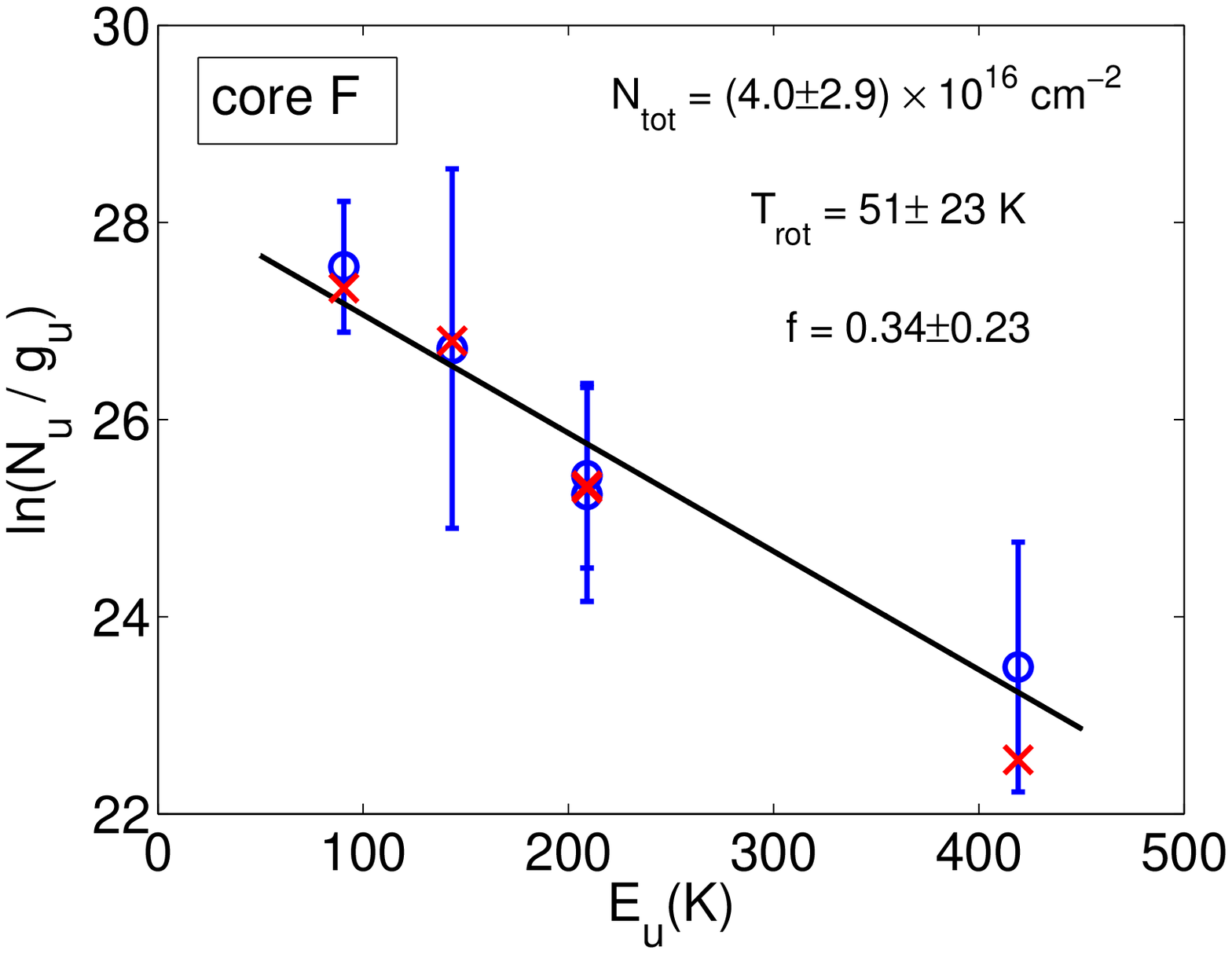}
\end{minipage}
\begin{minipage}[c]{0.5\textwidth}
  \centering
  \includegraphics[width=80mm,height=70mm,angle=0]{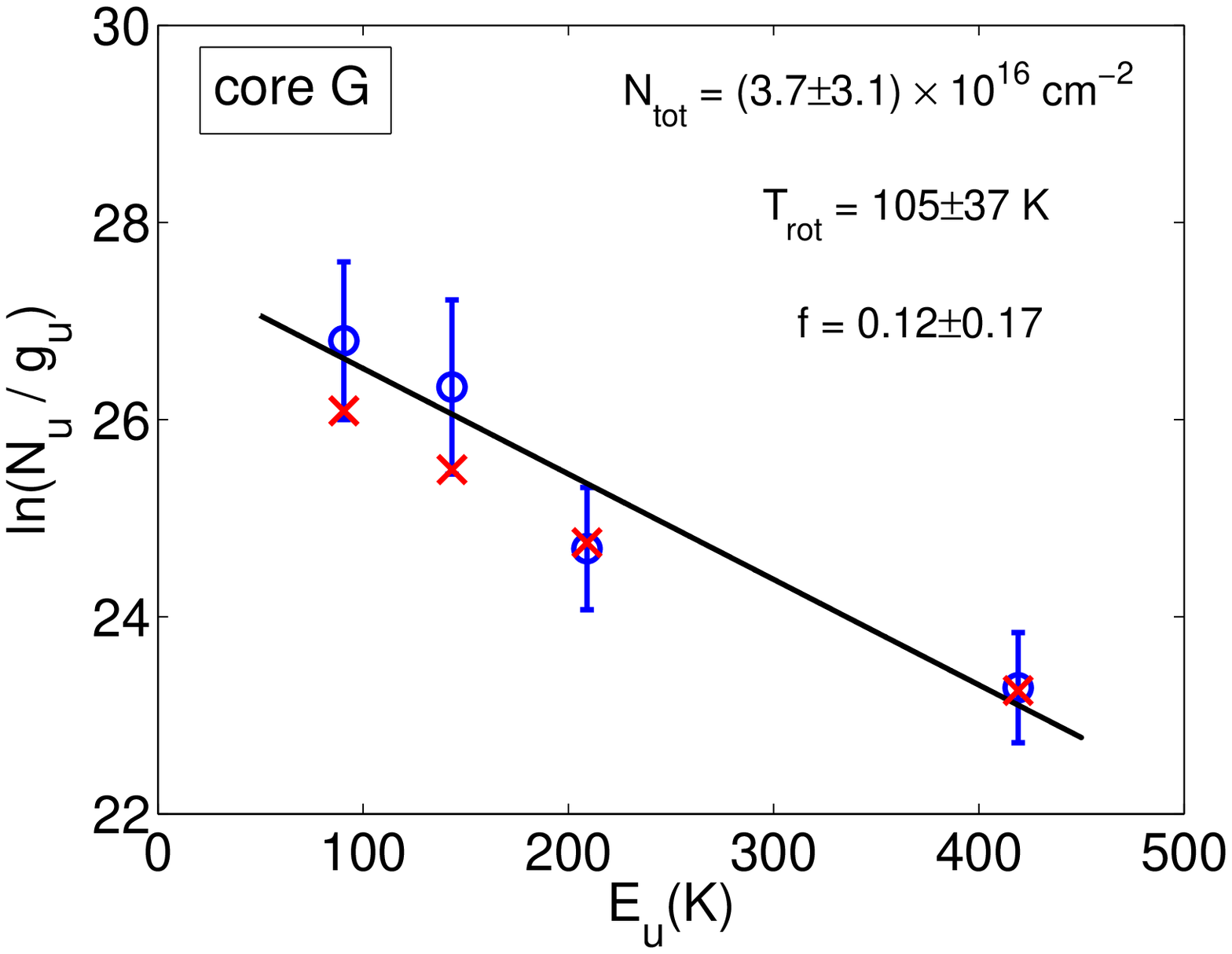}
\end{minipage}
\caption{Population diagrams of H$_{2}$CS towards four cm/mm cores.
The names of the cores are labeled on the upper-right corner of each
panel. Open circles in blue represent the observed data. The
vertical bars present 3$\sigma$ errors of ln(N$_{u}$/g$_{u}$) due to
the uncertainties of integrated intensities. The solid line shows
the linear least-squares fitting using the Rotational Temperature
Diagram method. Crosses in red mark the weighted mean results from
Population Diagram analysis. The inferred parameters from the
Population Diagram analysis are presented on the upper-right corners
of each panel.}
\end{figure}

\begin{figure}
\begin{minipage}[c]{0.5\textwidth}
  \centering
  \includegraphics[width=80mm,height=70mm,angle=0]{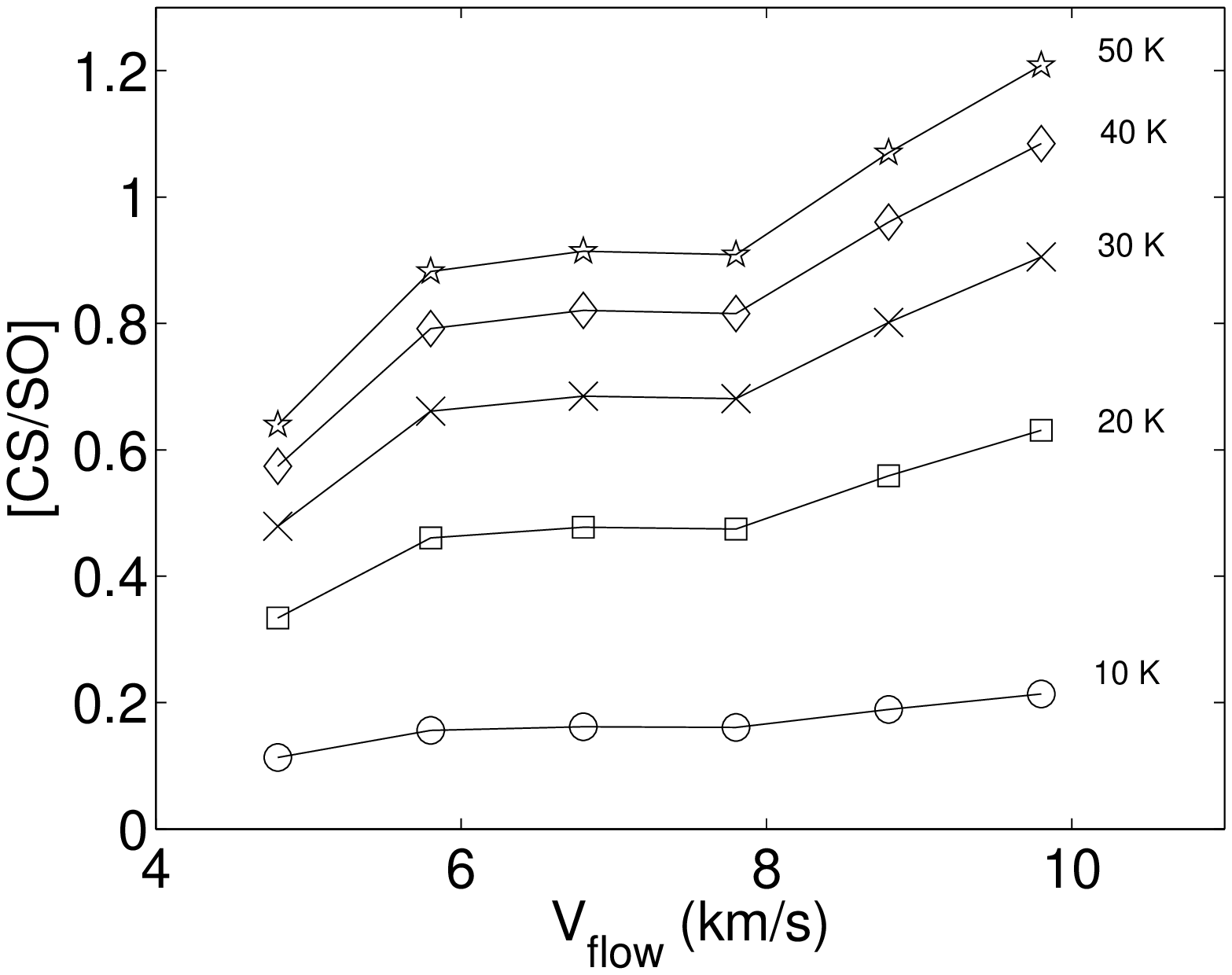}
\end{minipage}
\begin{minipage}[c]{0.5\textwidth}
  \centering
  \includegraphics[width=80mm,height=70mm,angle=0]{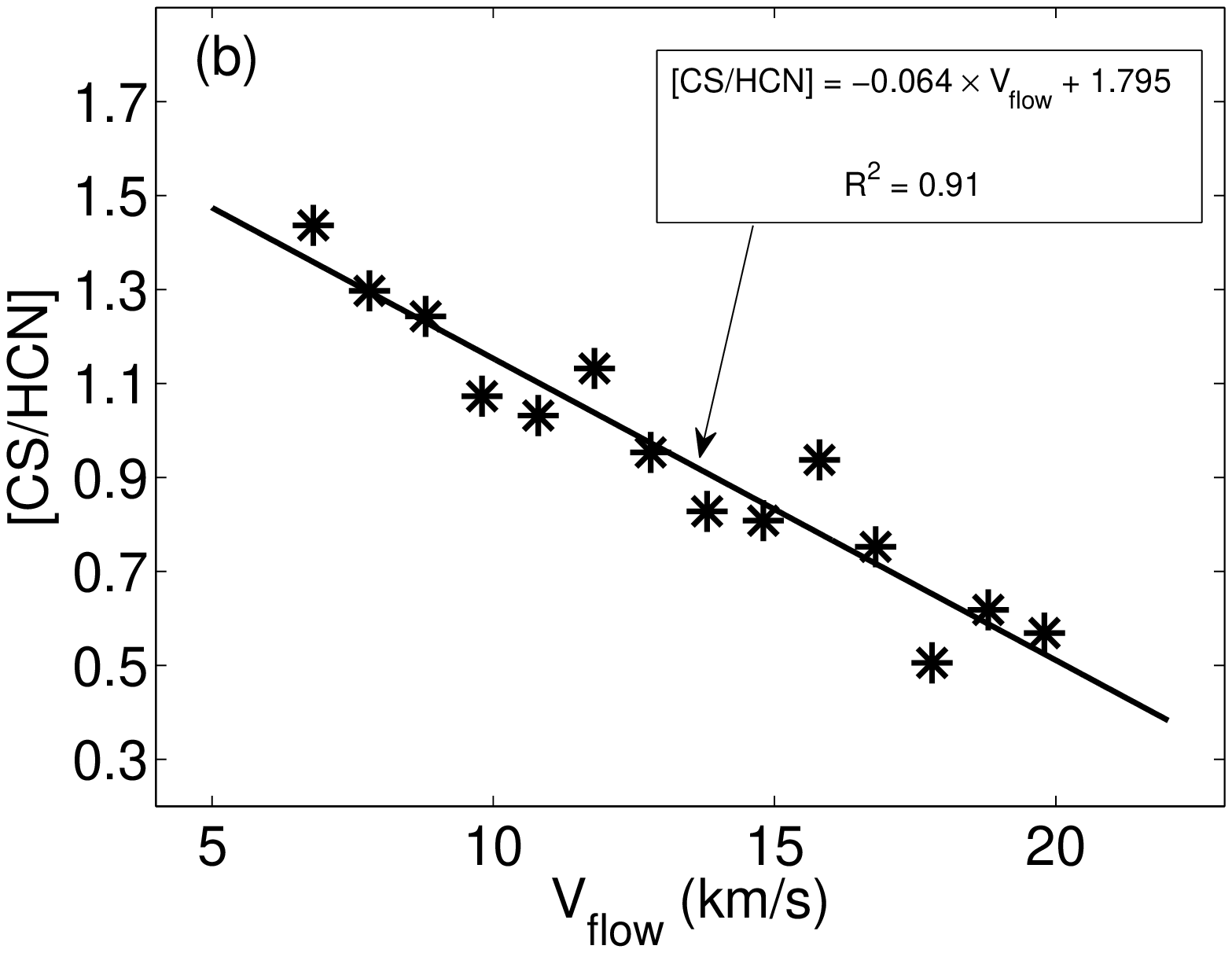}
\end{minipage}
\caption{Abundance ratios of [CS/SO] (left) and [CS/HCN] (right)
versus flow velocity along the redshifted lobe. We range the
excitation temperature from 10 K to 50 K to derive the abundance
ratios of [CS/SO]. The excitation temperature in calculation of
abundance ratios of [CS/HCN] is assumed to be 30 K. The solid line
in the right panel is the linear least-squares fitting, and the
fitting results are presented in the upper-right corner.}
\end{figure}

\begin{figure}
\begin{minipage}[c]{0.5\textwidth}
  \centering
  \includegraphics[width=80mm,height=70mm,angle=0]{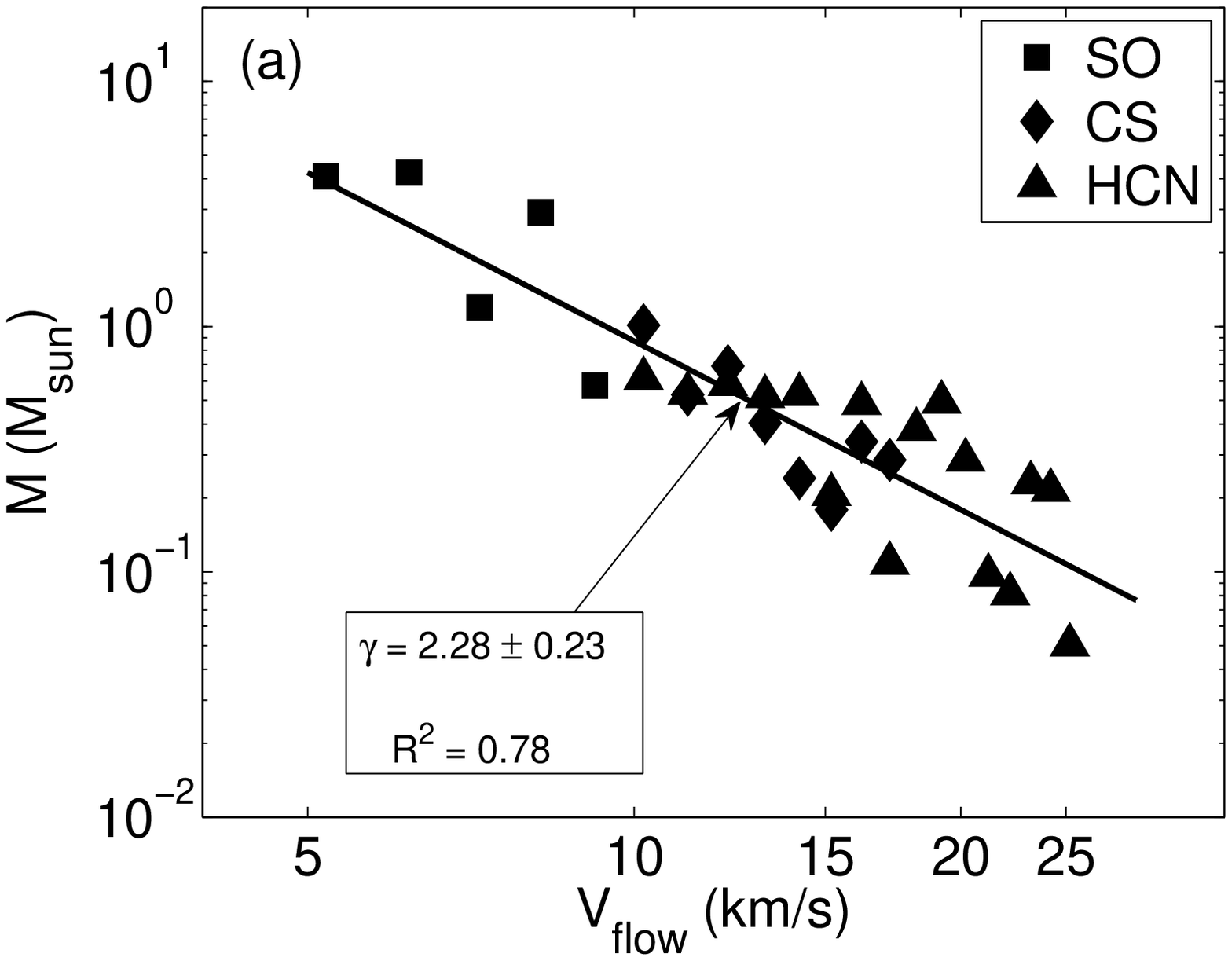}
\end{minipage}
\begin{minipage}[c]{0.5\textwidth}
  \centering
  \includegraphics[width=80mm,height=70mm,angle=0]{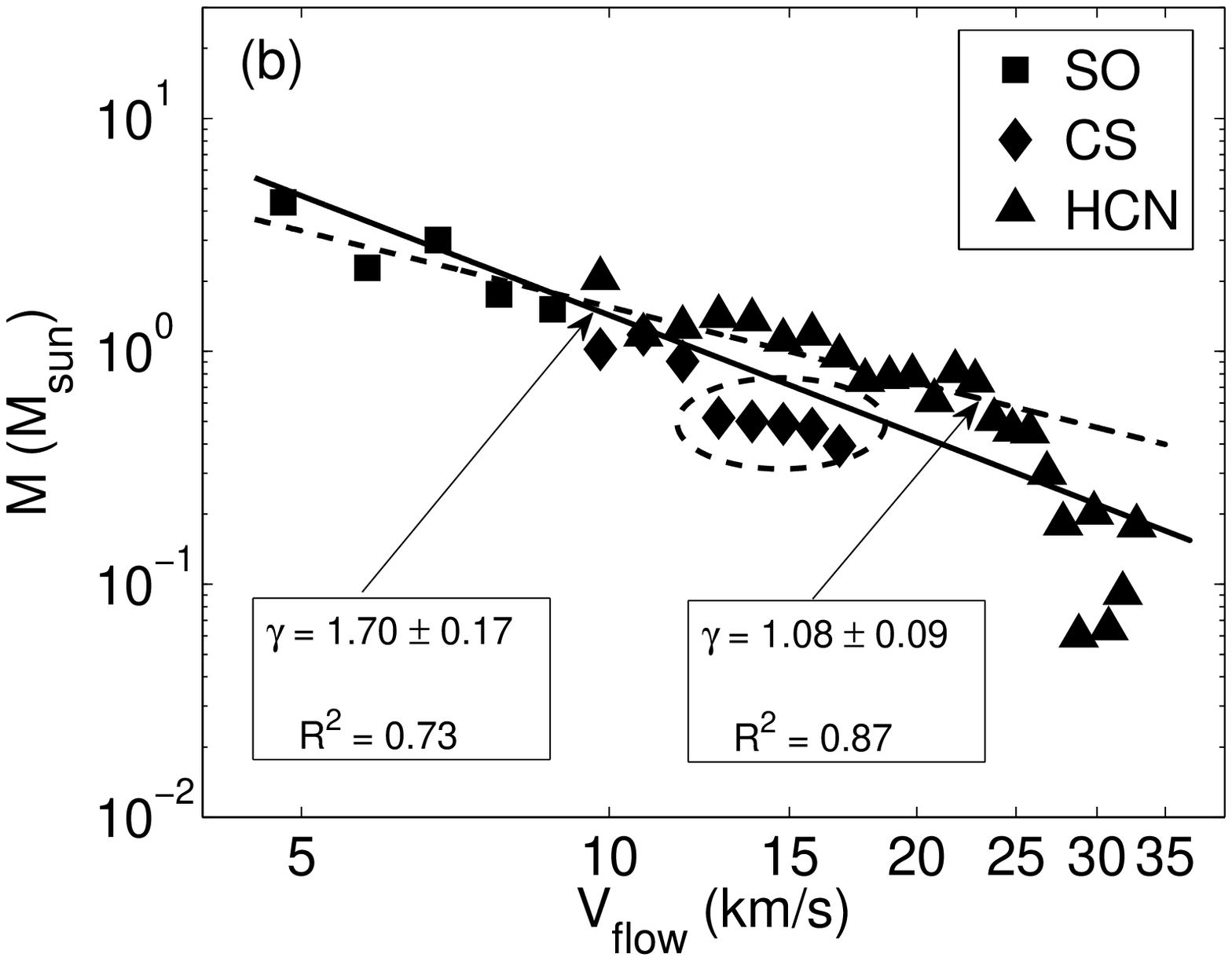}
\end{minipage}
\caption{Mass-Velocity relationships for the outflow lobes. Left :
blueshifted lobe; right: redshifted lobe.  The solid lines in both
panels show the power law fit towards all the data. The dashed line
in the right panel shows the power law fit towards the HCN and SO
data up to V$_{flow}$~=~25~km~s$^{-1}$. The fitting results are
presented in the lower-left corners.}
\end{figure}

\clearpage
\begin{deluxetable}{ccrrrrrrrrrcrl}
\tabletypesize{\scriptsize}  \tablecaption{Parameters of 860
$\micron$ continuum emission}
 \tablewidth{0pt} \tablehead{
  & R.A. & Decl. &  Deconvolution sizes & I$_{peak}$ & S$_{\nu}$  & T$_{d}$\tablenotemark{a} & $\beta$\tablenotemark{a} & Mass & N$_{H_{2}}$\\
   Name & (J2000)& (J2000) & ($\arcsec~\times~\arcsec$)&  (Jy~beam$^{-1}$) & (Jy)& (K) &
     &(M$_{\sun}$) & ($10^{24}$~cm$^{-2}$)
   }
\startdata

Northern core & 18:06:14.447 & -20:31:28.253 & Point source &
0.20$\pm$0.02 & 0.26 &
50 & 1.5 & 13&\\
Middle core & 18:06:14.668 & -20:31:31.830 &
$1.48\arcsec\times1.29\arcsec$ (P.A.=$-37.8\arcdeg$) & 0.76$\pm$0.04
&1.07 & 92 & 1.2 & 30 &1.2
\\
Southern core &  18:06:14.889 & -20:31:40.149 &
$4.71\arcsec\times1.26\arcsec$ (P.A.=$-20.2\arcdeg$) & 0.95$\pm$0.12
&2.52 & 51 &
0.8 & 165 &2.1\\
\enddata
\tablenotetext{a}{The dust temperature is assumed to be the same as
the rotational temperature of H$_{2}$CS transitions}
\tablenotetext{b}{The opacity index $\beta$ is obtained from
\cite{su05}}
\end{deluxetable}

\begin{deluxetable}{ccrrrrrrrrrrrrrrrrrrrrrrrrrcrl}
\rotate \tabletypesize{\scriptsize} \setlength{\tabcolsep}{0.05in}
\tablecolumns{16} \tablewidth{0pc} \tablecaption{Observed parameters
of the lines} \tablehead{ \colhead{Molecule}    &
\colhead{Transition} & \colhead{Frequency} &\colhead{E$_{u}$} &
\colhead{rms} & \multicolumn{4}{c}{V$_{lsr}$\tablenotemark{b}} &
\multicolumn{4}{c}{Intensity\tablenotemark{b}} &
\multicolumn{4}{c}{FWHM\tablenotemark{b}}  \\
\colhead{}    & \colhead{} & \colhead{(GHz)} & \colhead{(K)} &
\colhead{(Jy~beam$^{-1}$)} & \multicolumn{4}{c}{(km~s$^{-1}$)} &
\multicolumn{4}{c}{(Jy~beam$^{-1}$)} &
\multicolumn{4}{c}{(km~s$^{-1}$)}  \\
\cline{6-17} \\
\colhead{} & \colhead{}   & \colhead{}  & \colhead{} & \colhead{} &
\colhead{D} & \colhead{E} & \colhead{F}   & \colhead{G} &
\colhead{D} & \colhead{E} & \colhead{F}   & \colhead{G}    &
\colhead{D} & \colhead{E} & \colhead{F}   & \colhead{G}    }
\startdata
H$_{2}$CS          &10$_{0,10}$-9$_{0,9}$          &342.946  & 90.6 &0.3  & 5.5$\pm$0.2 &2.7$\pm$0.2  &5.9$\pm$0.2 &6.0$\pm$0.5 & 1.9$\pm$0.3 &1.4$\pm$0.2  &1.7$\pm$0.1  & 0.9$\pm$0.2 & 2.7$\pm$0.4 & 3.1$\pm$0.5 &4.3$\pm$0.6  &  3.8$\pm$1.1   \\
                   &10$_{2,9}$-9$_{2,8}$\tablenotemark{c}           &343.322 &143.3 &0.3  & 4.0$\pm$0.5 &2.2$\pm$1.1  &            &4.9$\pm$0.3 & 0.9$\pm$0.2 &1.4$\pm$0.3  &             & 1.1$\pm$0.3 & 4.0$\pm$0.6 & 4.4$\pm$1.6 &             &  1.9$\pm$0.6   \\
                   &10$_{2,8}$-9$_{2,7}$           &343.813  &143.3 &0.3  & 6.0$\pm$0.3 &2.5$\pm$0.4  &4.9$\pm$0.5 &            & 1.3$\pm$0.2 &0.9$\pm$0.1  &0.8$\pm$0.2  &             & 3.0$\pm$0.7 & 5.5$\pm$0.9 &3.9$\pm$1.3  &                \\
                   &10$_{3,8}$-9$_{3,7}$\tablenotemark{d}           &343.410&209.1  &0.3  & 6.2$\pm$0.8 &2.4$\pm$0.7  &5.0$\pm$0.3 &            & 1.1$\pm$0.7 &1.1$\pm$0.1  &2.1$\pm$0.3  &             & 3.3$\pm$0.7 & 4.1$\pm$0.5 &2.7$\pm$0.6  &                \\
                   &10$_{3,7}$-9$_{3,6}$\tablenotemark{e}           &343.414 &209.1 &0.3  & 5.8$\pm$2.8 &2.6$\pm$0.2  &5.5$\pm$0.3 &5.8$\pm$0.2 & 0.7$\pm$0.2 &2.0$\pm$0.2  &2.4$\pm$0.3  & 1.5$\pm$0.3 & 5.9$\pm$0.8 & 4.0$\pm$0.5 &2.9$\pm$0.6  &  2.3$\pm$0.5    \\
                   &10$_{5,6/5}$-9$_{5,5/4}$\tablenotemark{f}       &343.203  &419.2 &0.2  &             &1.8$\pm$0.4  &5.9$\pm$0.3 &5.9$\pm$0.2 &             &0.6$\pm$0.2  &0.7$\pm$0.2  & 0.9$\pm$0.2 &             & 3.2$\pm$1.0 &2.4$\pm$0.8  &  1.5$\pm$0.4 \\
SO                 & 8$_{8}$-7$_{7}$               & 344.311   &87.5 &0.2  & 4.9$\pm$0.1 &2.2$\pm$0.1  &5.0$\pm$0.1 &4.4$\pm$0.1 & 2.9$\pm$0.1 &4.0$\pm$0.1  &5.5$\pm$0.1  & 3.5$\pm$0.1 & 3.9$\pm$0.2 & 5.3$\pm$0.2 &9.0$\pm$0.2  &  8.2$\pm$0.3     \\
HC$^{15}$N~$\nu$=0 & 4-3                           & 344.200   &41.3 &0.2  & 6.1$\pm$0.4 &2.6$\pm$0.1  &5.2$\pm$0.1 &4.8$\pm$0.3 & 0.6$\pm$0.1 &2.1$\pm$0.1  &2.8$\pm$0.1  & 1.1$\pm$0.1 & 3.6$\pm$1.0 & 4.8$\pm$0.3 &8.0$\pm$0.3  &  7.5$\pm$0.8      \\
CS                 & 7-6                          & 342.883 &65.8   &0.3  &             &             &            &            &            &             &             &             &             &             &             &         \\
HCN~$\nu$=0        & 4-3                         & 354.505  &42.5  &0.3  &  &             &            & & &             &             &  &  &             &             &         \\
\enddata
\tablenotetext{a}{Not all the detected lines are listed in this
table. The others will be presented in another paper.}
\tablenotetext{b}{The V$_{lsr}$, Intensity and FWHM of each
transition are derived from single gaussian fit towards the
beam-averaged spectra.} \tablenotetext{b}{Blended with
H$_{2}^{13}$CO (5$_{1,4}$-4$_{1,4}$) at 343.325713
GHz.}\tablenotetext{d}{Blended with H$_{2}$CS
(10$_{3,7}$-9$_{3,6}$)} \tablenotetext{e}{Blended with H$_{2}$CS
(10$_{3,8}$-9$_{3,7}$)} \tablenotetext{f}{The two transitions of
H$_{2}$CS (10$_{5,6}$-9$_{5,6}$) and (10$_{5,6}$-9$_{5,6}$) have the
same frequency, line strength and permanent dipole moment. Therefore
they has same contributions to the observed line profile.}
\end{deluxetable}

\begin{deluxetable}{ccrrrrrrrrrrrrrrrcrl} \rotate
\tabletypesize{\scriptsize} \tablecolumns{13} \tablewidth{0pc}
\tablecaption{The physical parameters of H$_{2}$CS transitions
obtained with Rotational Temperature Diagram (RTD) method and
Population Diagram (PD) analysis} \tablehead{
 \colhead{Core}& \multicolumn{2}{c}{RTD} & \colhead{}
&\multicolumn{9}{c}{PD}\\
\cline{2-3} & \cline{4-12} \\
\colhead{}  & \colhead{T$_{rot}$ (K)} & \colhead{N$_{tot}$
(10$^{15}$~cm$^{-2}$)} & \colhead{} & \colhead{T$_{rot}$ (K)} &
\colhead{N$_{tot}$ (10$^{16}$~cm$^{-2}$)} & \colhead{f} &
\multicolumn{6}{c}{$\tau$} \\
\cline{8-13} \\
\colhead{} & \colhead{}   & \colhead{}   & \colhead{} & \colhead{} &
\colhead{}& \colhead{} & \colhead{(10$_{0,10}$-9$_{0,9}$)}   &
\colhead{(10$_{2,9}$-9$_{2,8}$)} & \colhead{(10$_{2,8}$-9$_{2,7}$)}
& \colhead{(10$_{3,8}$-9$_{3,7}$)} &
\colhead{(10$_{3,7}$-9$_{3,6}$)} &
\colhead{(10$_{5,6/5}$-9$_{5,6/4}$)}} \startdata
D &43$\pm$9& 3.8$\pm$2.9& & 42$\pm$34 &4.2$\pm$2.9 & 0.46$\pm$0.24 & 6.4$\pm$4.4 & 0.7$\pm$0.5 & 0.9$\pm$0.6 & 0.4$\pm$0.5 & 0.2$\pm$0.3 & \\
E &83$\pm$21& 2.5$\pm$1.6& & 92$\pm$74 &3.6$\pm$3.0 & 0.26$\pm$0.23 & 4.1$\pm$4.3 & 0.7$\pm$0.7 & 0.6$\pm$0.6 & 0.7$\pm$0.8 & 0.7$\pm$0.8 &0.1$\pm$0.1\\
F &83$\pm$7& 2.6$\pm$0.6& & 51$\pm$23 &4.0$\pm$2.9 & 0.34$\pm$0.23 & 3.9$\pm$3.1 &             & 1.1$\pm$0.9 & 1.1$\pm$1.2 & 1.1$\pm$1.2 &0.0$\pm$0.1\\
G &91$\pm$17&1.3$\pm$0.7 & & 105$\pm$37 &3.7$\pm$3.1 & 0.12$\pm$0.18 & 2.5$\pm$2.7& 2.4$\pm$2.3 &             &             & 2.4$\pm$2.1 &0.3$\pm$0.3\\
\enddata
\tablenotetext{a}{The rotational temperature and total column
density of H$_{2}$CS transitions derived from RTD analysis are
presented in the second and third columns, while those derived from
PD analysis are shown in the forth and fifth columns. The sixth
column gives the filling factor of each source inferred from PD
analysis. The last six columns exhibit the optical depth of each
transition using PD analysis. }
\end{deluxetable}

\begin{deluxetable}{ccrrrrrrrrrrrrrcrl}
\tabletypesize{\scriptsize} \tablecolumns{15} \tablewidth{0pc}
\tablecaption{Outflow parameters of the southern core} \tablehead{
\colhead{Molecule} & \multicolumn{2}{c}{Velocity interval} &
\multicolumn{2}{c}{M} &
\multicolumn{2}{c}{P} & \multicolumn{2}{c}{E} \\
\colhead{} & \multicolumn{2}{c}{(km~s$^{-1}$)} &
\multicolumn{2}{c}{(M$_{\sun}$)} &
\multicolumn{2}{c}{(M$_{\sun}~\cdot~km~s^{-1}$)} & \multicolumn{2}{c}{($10^{45}$erg)}  \\
\cline{2-9} \\
\colhead{Component} & \colhead{Blue}   & \colhead{Red}    &
\colhead{Blue} & \colhead{Red} & \colhead{Blue}   & \colhead{red} &
\colhead{Blue} & \colhead{Red} } \startdata SO
&[-4,0]&[10,14]&13&13&86&82&5.8&5.4\\
CS&[-12,-5]&[15,22]&3.7&5.5&47&68&6.0&8.7\\
HCN&[-20,-5]&[15,39]&5.4&17.6&85&294&14.1&54.6\\
\enddata
\end{deluxetable}


\clearpage
\begin{thebibliography}{}
{\small
\bibitem[Arce et al.(2007)]{arc07}Arce, H. G., Shepherd, D., Gueth, F., Lee, C.-F., Bachiller, R., Rosen, A., Beuther, H., 2007, Protostars and Planets
V, p. 245
\bibitem[Bachiller et al.(1997)]{bac97}Bachiller, R., Perez Guti\'{e}rrez, M. 1997, \apj, 487, L93
\bibitem[Blake et al.(1987)]{bl87}Blake, G. A., Sutton, E. C., Masson, C. R., \& Phillips, T. G., 1987, \apj, 315, 621
\bibitem[Bonnell et al.(1998)]{bon98}Bonnell, I. A., Bate, M. R., \& Zinnecker, H., 1998, \mnras, 298, 93
\bibitem[Cesaroni et al.(1994)]{ces94}Cesaroni, R., Churchwell, E., Hofner, P., Walmsley, C. M., \& Kurtz, S., 1994, \aap, 288, 903
\bibitem[Chandler et al.(1996)]{cha96}Chandler, C. J., Terebey, S., Barsony, M., Moore, T. J. T., \& Gautier, T. N., 1996, \apj, 471, 308
\bibitem[Choi(2002)]{cho02}Choi, M., 2002, \apj, 575, 900
\bibitem[Choi et al.(2004)]{cho04}Choi, M., Kamazaki, T., Tatematsu, K., \& Panis, J.-F., 2004, \apj, 617, 1157
\bibitem[Codella et al.(2005)]{cod05}Codella, C., Bachiller, R., Benedettini, M., Caselli, P., Viti, S., \& Wakelam, V., 2005, \mnras, 361, 244
\bibitem[Cummins, Linke, \&Thaddeus(1986)]{cum86}Cummins, S. E., Linke, R. A., Thaddeus, P., 1986, \apjs, 60, 819
\bibitem[Forster \& Caswell(1989)]{for89}Forster, J. R. \& Caswell, J. L., 1989, \aap, 213, 339
\bibitem[Fuller, Williams, \& Sridharan(2005)]{ful05}Fuller, G. A., Williams, S. J., \& Sridharan, T. K., 2005, \aap, 442, 949
\bibitem[Furuya, Cesaroni, \& Shinnaga(2011)]{fur11}Furuya, R. S., Cesaroni, R., \& Shinnaga, H., 2011, \aap, 525, 72
\bibitem[Garay et al.(1993)]{gar93}Garay, G., Rodriguez, L. F., Moran, J. M., \& Churchwell, E., 1993, \apj, 418, 368
\bibitem[Gibb, Wyrowski, \& Mundy (2004)]{gib04}Gibb, A. G., Wyrowski, F., \& Mundy, L. G., 2004, \apj, 616, 301
\bibitem[Goedhart, Gaylard, \& van der Walt(2005)]{goe05}Goedhart, S., Minier, V., Gaylard, M. J., \& van der Walt, D. J., 2005, \mnras, 356, 839
\bibitem[Goldsmith, \& Langer(1999)]{gol99}Goldsmith, P. F., \& Langer, W. D. 1999, \apj, 517, 209
\bibitem[Hildebrand (1983)]{hil83}Hildebrand, R. H. 1983, QJRAS, 24, 267
\bibitem[Hofner et al.(1994)]{hof94}Hofner, P., Kurtz, S., Churchwell, E., Walmsley, C. M., \& Cesaroni, R., 1994, \apj, 429, L85
\bibitem[Hofner, \& Churchwell(1996a)]{hof96a}Hofner, P., \& Churchwell, E., 1996a, \aaps, 120, 283
\bibitem[Hofner et al.(1996b)]{hof96b}Hofner, P., Kurtz, S., Churchwell, E., Walmsley, C. M., \& Cesaroni, R., 1996b, \apj, 460, 359
\bibitem[Hofner, Wiesemeyer, \& Henning (2001)]{hof01}Hofner, P., Wiesemeyer, H., \& Henning, T., 2001, \apj, 549, 425
\bibitem[Jiang et al.(2005)]{jia05}Jiang, Z., Tamura, M., Fukagawa, M., Hough, J., Lucas, P., Suto, H., Ishii, M., Yang, J., 2005, \nat, 437, 112
\bibitem[J\"{o}rgensen, Sch\"{o}ier, \& van Dishoeck(2004)]{jor04}J\"{o}rgensen, J. K., Hogerheijde, M. R., Blake, G. A., van Dishoeck, E. F., Mundy, L. G., Sch\"{o}ier, F. L., 2004, \aap, 415, 1021
\bibitem[Keto, Ho,\& Haschick(1988)]{ket88}Keto, E. R., Ho, P. T. P., \& Haschick, A. D., 1988, \apj. 324, 920
\bibitem[Keto(2002)]{keto02}Keto, E., 2002, \apj, 280, 580
\bibitem[Klaassen, \& Wilson(2007)]{kla07}Klaassen, P. D., \& Wilson, C. D., 2007, \apj, 663, 1092
\bibitem[Kurtz, \& Franco(2002)]{kur02}Kurtz, S., \& Franco, J., 2002, RMxAC, 12, 16
\bibitem[Lada, \& Fich(1996)]{lad96}Lada, C. J. \& Fich M., 1996, \apj, 459, 638
\bibitem[Lamers et al.(1995)]{lam95}Lamers, H. J. G. L. M., Snow, T. P., Lindholm, D. M. 1995, \apj, 455, 269
\bibitem[Linz et al.(2005)]{linz05}Linz, H., Stecklum, B., Henning, T., Hofner, P., Brandl, B., 2005, \aap, 429, 903
\bibitem[Liu, et al.(2002)]{liu02}Liu, S., Girart, J. M., Remijan, A., \& Snyder, L. E. 2002, \apj, 576, 255
\bibitem[Liu et al.(2011)]{liu11}Liu, T., Wu, Y., Zhang, Q., Ren, Z., Guan, X., \& Zhu,
M., 2011, \apj, 727, 1
\bibitem[Longmore et al.(2007)]{lon07}Longmore, S. N., Burton, M. G., Barnes, P. J., Wong, T., Purcell, C. R., \& Ott, J., 2007, IAUS, 242, 125
\bibitem[Mardones et al.(1997)]{mar97}Mardones, D., Myers, P. C., Tafalla, M., Wilner, D. J., Bachiller, R., \& Garay,
G., 1997, \apj, 489, 719
\bibitem[Nilsson et al.(2000)]{nil00}Nilsson, A., Hjalmarson, ${\AA}$., Bergman, P., Millar, T.
J., 2000, \aap, 358, 257
\bibitem[Norris et al.(1993)]{nor93}Norris, R. P., Whiteoak, J. B., Caswell, J. L., Wieringa, M. H., \& Gough, R. G., 1993, \apj, 412, 222
\bibitem[Ossenkopf \& Henning(1994)]{ossen94}Ossenkopf, V., Henning, T. 1994, \aap, 291, 943
\bibitem[Patel et al.(2005)]{pat05}Patel, N. A., et al., 2005, \nat, 437, 109
\bibitem[Persi et al.(2003)]{per03}Persi, P., Tapia, M., Roth, M., Marenzi, A. R., Testi, L., \& Vanzi, L., 2003, \aap, 397, 227
\bibitem[Qin et al.(2008a)]{qin08a}Qin, S., Wang, J., Zhao, G., Miller, M., \& Zhao, J., 2008a, \aap, 484, 361
\bibitem[Qin et al.(2008b)]{qin08b}Qin, S., et al., 2008b, \apj, 677, 353
\bibitem[Qin et al.(2008c)]{qin08c}Qin, S.-L., Huang, M., Wu, Y., Xue, R., Chen, S., 2008c, \apj, 686, L21
\bibitem[Qin et al.(2010)]{qin10}Qin, S.-L., Wu, Y., Huang, M., Zhao, G., Li, D., Wang, J.-J., Chen,
S., 2010, \apj, 711, 399
\bibitem[Qiu et al.(2007)]{qiu07}Qiu, K., Zhang, Q., Beuther, H., \& Yang, J., 2007, \apj, 654, 361
\bibitem[Qiu et al.(2008)]{qiu08}Qiu, Keping., 2008, \apj, 685, 1005
\bibitem[Qiu et al.(2009)]{qiu09}Qiu, K., Zhang, Q., Wu, J., \& Chen, H.-R., 2009, \apj, 696, 66
\bibitem[Remijan et al.(2004)]{re04}Remijan, A., Sutton, E. C., Snyder, L. E., Friedel, D. N., Liu, S.-Y., \& Pei, C.-C., 2004, \apj, 606, 917
\bibitem[Ridge, \& Moore(2001)]{rid01}Ridge, N. A., \& Moore, T. J. T., 2001, \aap, 378, 495
\bibitem[Sault et al.(1995)]{sau95}Sault, R. J., Teuben, P. J., \& Wright, M. C. H. 1995, in ASP Conf. Ser. 77, Astronomical Data Analysis Software and Systems IV, ed. R. A. Shaw,
H. E. Payne, \& J. J. E. Hayes (San Francisco, CA: ASP), 433
\bibitem[Shu, Adams \& Lizano(1987)]{shu87} Shu, F. H., Adams, F. C., \& Lizano, S., 1987, \araa, 25, 23
\bibitem[Sridharan, Williams, \& Fuller(2005)]{sri05}Sridharan, T. K., Williams, S. J., \& Fuller, G. A., 2005, \apj, 631, L73
\bibitem[Su, Zhang, \& Lim(2004)]{su04}Su, Y.-N., Zhang, Q., \& Lim, J., 2004, \apj, 604, 258
\bibitem[Su et al.(2005)]{su05}Su Y.-N., Liu S.-Y., Lim J., Chen H.-R., 2005, in Protostars and Planets V Submillimeter Observations of the High- Mass Star
Forming Complex G9.62+0.19. pp 8336$-+$
\bibitem[Su et al.(2007)]{su07}Su, Y-N., Liu, S.-Y.., Chen, H.-R., Zhang, Q., Cesaroni, R., 2007, \apj, 671,571
\bibitem[Sun, \& Gao(2008)]{sun08}Sun, Y., \& Gao, Y., 2008, \mnras, 392, 170
\bibitem[Takami et al.(2010)]{tak10}Takami, M., Karr, J. L., Koh, H., Chen, H.-H., Lee, H.-T., 2010, \apj, 720, 155
\bibitem[Testi et al.(1998)]{tes98}Testi, L., Felli, M., Persi, P., \& Roth, M., 1998, \aap, 329, 233
\bibitem[Testi et al.(2000)]{tes00}Testi, L., Hofner, P., Kurtz, S., \& Rupen, M., 2000, \aap, 359, L5
\bibitem[Turner et al.(1991)]{tur91}Turner B. E., 1991, \apjs, 76, 617
\bibitem[van der Walt, Goedhart, \& Gaylard(2009)]{van09}van der Walt, D. J., Goedhart, S., \& Gaylard, M. J., 2009, \mnras, 298, 961
\bibitem[Van der Tak \& Menten(2005)]{van05}Van der Tak, F. F. S., \& Menten, K. M., 2005, \aap, 437, 947
\bibitem[Wang et al.(2010)]{wang10}Wang, K.-S., Kuan, Y.-J., Liu, S.-Y., Charnley, S. B., 2010, \apj, 713, 1192
\bibitem[Wolfire, \& Cassinelli(1987)]{wol87}Wolfire, M. G., \& Cassinelli, J. P., 1987, \apj, 319, 850
\bibitem[Wu \& Evans(2003)]{wu03}Wu, J., \& Evans N. J. II., 2003, \apj, 592, L79
\bibitem[Wu et al.(2004)]{wu04}Wu, Y., Wei, Y., Zhao, M., Shi, Y., Yu, W., Qin, S., Huang, M., 2004, \aap, 426, 503
\bibitem[Wu et al.(2005)]{wu05}Wu, Y., Zhu, M., Wei, Y., Xu, D., Zhang, Q., \& Fiege, J. D., 2005, \apj, 628, L57
\bibitem[Wu et al.(2007)]{wu07}Wu, Y., Henkel, C., Xue, R., Guan, X., \& Miller, M., 2007, \apj, 669, L37
\bibitem[Wu et al.(2009)]{wu09}Wu, Y., Qin, S.-L., Guan, X., Xue, R., Ren, Z., Liu, T., Huang, M., Chen,
S., 2009, \apj, 697, L116
\bibitem[Wyrowski et al.(2006)]{wyr06}Wyrowski, F., Heyminck, S., Gsten, R., \& Menten, K. M., 2006, \aap, 454, L95
\bibitem[Wyrowski(2007)]{wyr07}Wyrowski, F., 2007, ASPC, 387, 3W
\bibitem[Zhang, Ho, \& Ohashi(1998)]{zha98}Zhang, Q., Ho, P. T. P., \& Ohashi, N., 1998, \apj, 494, 636
\bibitem[Zhang et al.(2005)]{zha05}Zhang, Q, Hunter, T. R., Brand, J., Sridharan, T. K., Cesaroni, R., Molinari, S., Wang, J., Kramer, M., 2005, \apj, 625, 864
\bibitem[Zhou et al.(1993)]{zho93}Zhou, S., Evans, N. J. II., Koempe, C., \& Walmsley, C. M., 1993, \apj, 404, 232
\bibitem[Zinnecker \& Yorke(2007)]{zin07}Zinnecker, H., \& Yorke, H. W., 2007, \araa, 45, 481}
\end{thebibliography}
\end{document}